# Tailoring the Phase Transition and Electron-Phonon Coupling in 1T′-MoTe$_2$ by Charge Doping: A Raman Study

Suvodeep Paul, Saheb Karak, Manasi Mandal, Ankita Ram, Sourav Marik, R. P. Singh, and Surajit Saha*

(* surajit@iiserb.ac.in)

*Department of Physics, Indian Institute of Science Education and Research Bhopal, Bhopal, 462066, India*

**Abstract**

Transition metal dichalcogenides (TMDs) are a class of widely studied 2D layered materials which exist in various polymorphs. The 1T′ phase of MoTe$_2$ is of prime importance as it has been reported to show quantum spin hall (QSH) behavior with a fairly large band-gap of ~ 60 meV, in contrast to most QSH materials known. It is noteworthy that though the monolayer 1T′-MoTe$_2$ was initially predicted to show the QSH behavior, recent theoretical studies claim that the few-layered counterparts also exhibit higher order topological behavior. Besides, 1T′-MoTe$_2$ also undergoes a hysteretic phase transition to the T$_d$ phase (which is a type-II Weyl semimetal) by breaking the inversion symmetry of the crystal. While the phase transition between these two topological phases is of utmost importance, its study has been mostly restricted to bulk single crystal flakes, thereby not sufficiently exploring the effect of dimensionality. We have studied the phase transition in 1T′-MoTe$_2$ as a function of flake-thickness. Though our Raman studies show a suppression of the phase transition in the thin (thickness <10 nm) flakes [similar to the report Phys. Rev. B **97**, 041410 (2018)], we have experimentally demonstrated the possibility of stabilizing the desired phase (1T′ or T$_d$) at room temperature by charge doping. Further, we have observed clear signatures of electron-phonon coupling in MoTe$_2$, which evolves as a function of flake-thickness and charge doping.

## 1. Introduction

Since the inception of the field of 2D materials with the discovery of graphene by Geim and Novoselov [1], scientists have explored a variety of layered materials [2]. The weak out-of-plane van-derWaals interaction made it possible to exfoliate dangling bonds-free single layers of theses materials, thereby allowing the study of various quantum phenomena in lower dimensions [3]. The transition metal dichalcogenides (TMDs), which exist in a variety of polymorphs [4,5], are a class of widely studied 2D materials. While the semiconducting 2H phase has been demonstrated to show many interesting properties and potential applications [6-8], the semimetallic phases are still relatively new to the scientific community. Recent investigations of these phases have predicted and demonstrated various important topological phenomena in TMDs such as Quantum spin Hall effect (QSH) and presence of Weyl fermions. Through first principles calculations, Qian *et al*. [9] proposed the monolayers of 1T′ TMDs to show QSH effect. QSH phases are characterized by topologically protected helical edge states and are manifested by 2D systems showing intrinsic band inversion and band gap opening due to strong spin-orbit coupling [9,12]. A recent theoretical report by Wang *et al.* [13], however, claims that multi-layer counterparts of 1T′-$MoTe_2$ also exhibit higher order topological behavior. The structural distortion of the 1T phase into 1T′ has been theoretically established to cause the band inversion [9,12,14]. Recently, Angle-Resolved Photo-Emission Spectroscopy (ARPES) measurements on 1T′-$WTe_2$ revealed the opening of a bandgap, confirming its QSH phase [10]. Keum *et al*. [12] experimentally demonstrated the bandgap opening of ~ 60 meV to take place in few-layered 1T′-$MoTe_2$. QSH materials promise great potential in spintronic applications, however, the low bandgap in most of the explored QSH materials [15-17] limits their spintronic applications to low temperatures. The bandgap opening reported in 1T′-$MoTe_2$ is sufficiently large, thus enabling the possibility for high temperature spintronic applications.

Charge transport measurements on bulk 1T′-$MoTe_2$ single crystals [18] revealed a hysteretic phase transition below ~ 250 K into the $T_d$ phase. The $T_d$ phase of $WTe_2$ and $MoTe_2$ has been predicted to be Type-II Weyl semimetals (WSM) [11], which are platforms for the realization of Weyl fermions (predicted in high energy particle physics) in condensed matter systems [19-29]. Naturally, switching between the two topological phases (1T′ and $T_d$) is of great significance. According to He *et al*. [30], all semimetallic $MoTe_2$ flakes having thicknesses below a critical value ~ 20 nm exist in the $T_d$ phase. There is a recent report on the observation of the $T_d$ phase at room temperature for molecular beam epitaxy (MBE) grown few-layered $MoTe_2$ [31]. The complete suppression of the QSH phase contradicts the reports by Qian *et al*. [9] and Keum *et al*. [12] thus lessening the importance of $MoTe_2$ as an excellent topological material. A recent report demonstrates the stabilization of the 1T′ phase of $WTe_2$ by inducing ultrafast interlayer sheer strain using terahertz light pulses [32]. On the other hand, Kim *et al*. [33] theoretically predicted the stabilization of the 1T′ phase by hole doping, while electron doping was predicted to stabilize the $T_d$ phase. Motivated by these predictions, we have performed experiments to verify the possibility of switching between the 1T′ and $T_d$ phases at room temperature. To the best of our knowledge, switching of 1T′ / $T_d$ phase by hole / electron doping has not yet been experimentally demonstrated. However, there has been a recent report on subpicosecond optical switching between the topological phases in bulk MoTe$_2$ by using pumb-probe spectroscopy [34].

The inherent characteristic requirement of WSMs is that the system breaks time reversal symmetry or lattice inversion symmetry [11]. The time-reversal symmetry conserving WSM systems like WTe$_2$ and MoTe$_2$ must break the lattice inversion symmetry of the 1T′ crystal to attain the Weyl semimetallic $T_d$ phase. The breaking of lattice inversion symmetry can be very well probed by Raman spectroscopy, which is a simple, non-destructive characterizing tool for all 2D materials. The WSM phase in MoTe$_2$ has been confirmed using Raman spectroscopy by various groups where a new phonon mode appears in the WSM ($T_d$) phase as compared to the 1T′ phase [35-39]. Importantly, MoTe$_2$ has also been reported to show superconductivity with a $T_c$ of ~ 0.1 K [40]. The origin of superconductivity has often been attributed to electron-phonon coupling, which is generally manifested as phonon anomalies that have been extensively studied theoretically and experimentally for various superconductors [41,42]. Again, Raman is a very important technique for the detection of electron-phonon coupling in condensed matter systems [43-46]. Notably, most of the reports on the phase transition in MoTe$_2$ have been limited to bulk flakes. So, a systematic study of the phase transition as a function of varying flake thickness is of great importance. In our work, we have performed extensive temperature-dependent Raman studies on flakes of various thicknesses to investigate the phase transition between the two topological phases of semimetallic MoTe$_2$ and the effect of charge doping on the phase transition. We demonstrate that hole (or electron) doping can stabilize the 1T′ (or $T_d$) phase at room temperature for MoTe$_2$ flakes of any arbitrary thickness thus making it even more profound a material with potentially switchable topological phases at room temperature. We have also observed signatures of electron-phonon coupling that evolves as a function of thickness of MoTe$_2$ and charge doping. Our detailed analysis of Raman data for flakes of MoTe$_2$ of various thicknesses down to monolayer brings forward several important results like complete renormalization of the phonons due the inversion symmetry breaking, tunability of phase transition temperature, stabilization of the desired topological phase (1T′ or $T_d$) at room temperature by charge doping in thin flakes, and the tunability of electron-phonon coupling with thickness.

## 2. Experimental details

Single crystals of 1T′-MoTe$_2$ were grown by Chemical Vapor Transport (CVT) method using iodine as the transport agent. In the first step, stoichiometric mixtures of Mo (99.9% pure), [Re (99.9% pure) for Re doped samples], and Te (99.99% pure) powders were ground together, pelletized, and sealed in an evacuated quartz tube. The sealed ampoule was first heated at 1100 °C for 24 hours, followed by ice water quenching to avoid formation of the 2H phase and the same heat treatment was repeated. The obtained phase-pure polycrystalline sample was used in the crystal growth process. Crystallization was carried out from hot zone (1100 °C) to cold zone (900 °C) in a two-zone furnace where the polycrystalline sample was kept in the hot zone. Finally, the sealed quartz tube was quenched in air to avoid the formation of the hexagonal phase. The phase formation was confirmed by single crystal X-ray diffraction (XRD) measurements (see Supplemental material [47] Fig. S1) performed using PANalytical diffractometer equipped with Cu-K$_\alpha$ radiation. Charge transport measurements were performed using a Quantum Design Physical Property Measurement System (PPMS). To study the effect of the flake thickness, bulk flakes were micromechanically exfoliated using scotch tape and then the obtained layers of 1T′-MoTe$_2$ were transferred onto silicon substrates coated with ~ 300 nm SiO$_2$ film. The Raman spectra were obtained in a Horiba JY LabRam HR Evolution Raman Spectrometer mounted with a grating of 1800 grooves/mm using a 50× objective (N.A. = 0.5) lens in back-scattering geometry. The samples were excited by a diode laser of wavelength 532 nm and the spot radius was ~ 0.7 μm. The detection was done by an air-cooled charge coupled device (CCD) detector. Temperature-dependent Raman measurements were performed in a LINKAM stage by varying the temperature from 80 K to 400 K. To avoid the effects of local heating by the laser, all measurements were performed using low power (~ 1 mW). Exfoliated samples of pure MoTe$_2$ were exposed to ambient atmosphere and ammonia gas to dope with holes and electrons, respectively. Additionally, Re incorporated MoTe$_2$, that induces electron doping in the layers, were also exfoliated down to monolayers to investigate the effects of charge doping on the 1T′ to $T_d$ phase transition and electron-phonon interactions.

## 3. Results

### 3.1 Phase transition in bulk 1T'-MoTe$_2$

At room temperature, bulk semimetallic MoTe$_2$ exists in the 1T′ phase. The structure of the 1T′ phase can be derived by introducing a distortion in the 1T phase (left column of Figure 1(a)) where the central layer of Mo atoms in the Te-Mo-Te units shows octahedral coordination with the Te atoms. The 1T phase is energetically unstable for MoTe$_2$ due to the phenomenon of fermi surface nesting and hence it undergoes a Peierls distortion thus dimerizing the Mo atoms [12] as shown in Figure 1(a). The resulting 1T′-MoTe$_2$ is monoclinic with point group $C^2_{2h}$ and space group $P2_1/m$. Figure 1(b) shows the crystal structure of 1T′ phase of bulk MoTe$_2$ that possesses inversion symmetry. Another stable phase can be formed by a slight modification in the stacking order of the Te-Mo-Te trilayers with respect to each other such that the resulting T$_d$ phase (Figure 1(c)) is orthorhombic with space group $Pnm2_1$. It can be clearly seen that the T$_d$ phase lacks inversion symmetry.

We have measured the charge transport in thick flakes of MoTe$_2$ with varying temperature. Figure 2(a) shows the resistivity as a function of temperature clearly revealing a hysteretic phase transition near 250 K matching with previous reports [18,35,37]. Figure 2(b) shows a comparison between the Raman spectra of 1T′-MoTe$_2$ taken at room temperature (300 K) and at 80K, respectively, where the phonon modes are labelled as $P_1$ to $P_9$ (see Supplemental material [47]). The mode $P_6$ clearly shows a splitting (to $P_{6A}$ and $P_{6B}$) in the spectrum obtained at 80 K. This is a very important observation in terms of the study of the phase transition in MoTe$_2$ from 1T′ to T$_d$ phase that undergoes an inversion symmetry breaking. The unit cell of bulk 1T′-MoTe$_2$ contains 12 atoms and hence there are 36 normal modes which reduce to the following irreducible representation at the Γ point of the Brillouin zone [35]: *Γ$_{bulk,1T'}$ = 12A$_g$+ 6A$_u$+ 6B$_g$+ 12B$_u$*. All the modes are either symmetric (*g*-type) or anti-symmetric (*u*-type) vibrations with respect to the centre of inversion symmetry. The *g* type modes are Raman active while the *u* type modes are Infra-red active. However, the low temperature phase (T$_d$) lacks inversion symmetry and, hence, all modes change from *A$_g$*, *B$_g$*, A$_u$, and *B$_u$* symmetry to *A$_1$*, *A$_2$*, *B$_1$*, and *B$_2$* types which are all Raman active (*A$_1$*, *B$_1$*, and *B$_2$* are also Infra-red active) [35]. As a result, the irreducible representation for the $T_d$ crystal is: $\Gamma_{bulk, Td} = 12A_1 + 6A_2 + 6B_1 + 12B_2$. Out of these, all phonons are Raman-active (except three acoustic modes of symmetries $A_1$, $B_1$, and $B_2$). There are *five* different modes (of Raman-inactive $A_u$ and $B_u$ symmetry) that convert to Raman active modes in the non-centrosymmetric phase, one of which appear as a low-frequency sheer mode near 11 cm$^{-1}$ (absent in our data due to optical filter cutoff) and another mode near 129 cm$^{-1}$ (which we labeled as $P_{6A}$ in the non-centrosymmetric phase). The other three predicted Raman modes could not be observed (by us and also by any group earlier), possibly due to a very low Raman cross-section. Therefore, the Raman spectrum shows a single mode ($P_6$) of $A_g$ symmetry near 129 cm$^{-1}$ in the 1T′ phase (as the other mode is a Raman silent $B_u$ mode) while the $T_d$ phase shows two modes ($P_{6A}$ and $P_{6B}$) in the region, both having $A_1$ symmetry. Therefore, the study of the phase transition would involve a detailed analysis of the evolution of mode $P_6$ into modes $P_{6A}$ and $P_{6B}$. Temperature-dependent Raman measurements near the phase transition temperature were performed over the heating and cooling cycles that show the phase transition at different temperatures indicating a hysteresis.

Figure 3(a) shows the evolution of the mode $P_6$ into the modes $P_{6A}$ and $P_{6B}$ in the heating and cooling cycles. The hysteretic character of the mode frequencies, linewidths, and integrated intensities of both the modes $P_{6A}$ and $P_{6B}$ are shown in Figure 3(b). This is expected as the breaking of inversion symmetry leads to a first order phase transition in 1T′-MoTe$_2$ which is accompanied by a complete renormalization of the phonons resulting in a hysteretic character of their parameters. The hysteresis shown by the linewidth of the mode $P_{6A}$ is poor as shown in Figure S2 in Supplemental material [47]. There are recent reports [35-37] which showed the hysteretic behavior of Raman intensity of the $P_{6A}$ mode due to the phase transition in 1T′-MoTe$_2$. However, the hysteresis observed in our samples are clearer and unlike the other reports which show the hysteresis in the intensity of the mode $P_{6A}$ alone; our data show a more complete picture of the renormalization of phonons resulting from the breaking of inversion symmetry.

*3.2 Tunability of the phase transition temperature*

In order to investigate the tunability of the phase transition temperature, we have doped the Mo sites in MoTe$_2$ by Re (Mo$_{1-x}$Re$_x$Te$_2$; x=0.1). Temperature-dependent Raman studies on Mo$_{1-x}$Re$_x$Te$_2$; x=0.1 (see Supplemental material [47] Figs S3 and S4) show a similar phase transition from the 1T′ to the $T_d$ phase. Hysteresis was observed in the phonon behavior of the modes $P_{6A}$ and $P_{6B}$. The hysteresis in intensities of modes $P_{6A}$ and $P_{6B}$ are compared with the hysteresis in pure MoTe$_2$ in Figure 4. It must be noted that Re doping shifts the centre of the hysteresis loops (labelled by solid and dashed lines for pure and Re doped MoTe$_2$, respectively) by ~ 30 K. Mandal *et al.* [48] reported that Re doping at the Mo sites causes electron doping in the system. We can, therefore, conclude that chemical (and charge) doping may be an effective way of tuning the phase transition temperature. Another interesting observation (see Supplemental material [47]) is the reversal in the behaviors of the P$_{6A}$ and P$_{6B}$ modes in Mo$_{1-x}$Re$_x$Te$_2$ with respect to their observed behaviors in pure MoTe$_2$. In the Re doped MoTe$_2$ samples, we observe that the P$_{6B}$ mode (and not the P$_{6A}$ mode) vanishes in the high temperature phase. The reason for reversal of this behavior is not clear at present. The incorporation of Re into the crystal is a complicated situation involving chemical (charge) doping along with local distortions in the crystal. Such local distortions may give rise to internal pressure/strain in the crystal which may in turn lead to other important phenomena like the elevation of the superconducting transition temperature [48] and appearance of pseudo magnetic field [49]. Therefore, a proper understanding of the phase transition pathways in Re doped MoTe$_2$ requires an in-depth study of the local distortion effects as well as the charge doping effects, both of which might lead to a modification of the phonon spectra. Such a study is beyond the scope of this article.

*3.3 Phonon anomaly in bulk 1T′-MoTe$_2$: Electron-phonon coupling*

Temperature-dependent Raman studies were performed on bulk (very thick) flakes of 1T′-MoTe$_2$ and the resulting spectra were analysed using Lorentzian multifunction. All the modes (marked as P$_2$ to P$_9$) show a redshift in mode frequency (see Supplemental material [47] Fig. S5) accompanied by an increase in linewidth (see Supplemental material [47] Fig. S6) with increasing temperature. These behaviors of the mode frequencies and linewidths can be explained by the anharmonic approximation for phonons (which also takes into account the thermal expansion contributions). The frequencies and linewidths follow the trends of the well-known cubic anharmonic equations [50-52]:

$$\omega_{ph}(T) = \omega_0 - C\left(1 + \frac{2}{e^{\frac{\hbar\omega_0}{2k_BT}} - 1}\right) \quad (1)$$

$$\Gamma_{ph}(T) = \Gamma_0 + \Gamma\left(1 + \frac{2}{e^{\frac{\hbar\omega_0}{2k_BT}} - 1}\right) \quad (2)$$

The typical behaviors of the mode frequency and linewidth as a function of temperature are shown in insets of Figure 5. In sharp contrast to the other phonons, the mode $P_1$ at ~ 78 cm$^{-1}$ shows a complete reversal of the trends in both frequency and linewidth as shown in Figures 5(a) and 5(b). In order to find an explanation to the anomalous behavior, we write the complete expression typically explaining the different factors that affect the frequency of a first-order Raman active phonon [43,46,53].

$$\omega(T) = \omega_0 + \Delta\omega_{vol}(T) + \Delta\omega_{anh}(T) + \Delta\omega_{sp-ph}(T) + \Delta\omega_{el-ph}(T) \tag{3}$$

This expression includes the quasi-harmonic contribution due to a change in unit cell volume ($\Delta\omega_{vol}$), the cubic anharmonicity effect due to phonon-phonon interactions ($\Delta\omega_{anh}$), spin-phonon coupling ($\Delta\omega_{sp-ph}$), and electron-phonon coupling ($\Delta\omega_{el-ph}$). The equation (1) takes care of the first three terms in the equation (3). As MoTe$_2$ is non-magnetic, contributions due to spin-phonon interactions can be completely ruled out. Therefore, we can attribute the anomalous temperature behavior of the $P_1$ phonon to electron-phonon coupling. It is to be noted that the presence of superconductivity [40] and charge density waves [48] have been reported in MoTe$_2$ systems. According to various reports, superconductivity is often induced by strong electron-phonon coupling which may be associated with anomalies in the phonon dispersion [41,42,43]. The phonons associated with such anomalies show anomalous temperature behavior in frequency (softening) and linewidth (broadening) due to smearing of the Fermi surface, leading to a reduction in electron-phonon coupling at higher temperatures [42]. Similar anomalous temperature behavior has also been witnessed in charge density wave (CDW) states formed due to Fermi-surface nesting. However, in absence of Fermi surface nesting, the softening of the phonon frequencies (and broadening of the corresponding linewidths) may be interpreted as a result of high degree of anhamonic nature of the bottom of the potential which gives rise to an increase in anharmonicity at lower temperatures. The electron-phonon coupling in graphene and graphite have been reported by studying the linewidth of the G band, which again show an anomalous behavior. Similar to other electron-phonon coupled systems, the anomalous temperature behavior of the $E_{2g}$ phonon (G band) linewidths in graphene and graphite has been attributed to dominance of electron-phonon coupling over its anharmonic nature [43,45,46].

Therefore, our attribution of the anomalous behavior of phonon frequency and linewidth of $P_1$ phonon mode in $MoTe_2$ to electron-phonon coupling is justified and this mode possibly plays an important role in the reported observation of superconductivity and charge density waves in 1T′-$MoTe_2$ systems. Additional evidence in support of electron-phonon coupling will be presented later.

*3.4 Thickness dependence of the phase transition of 1T′-MoTe$_2$*

The van der Waals interaction between the layers in 2D materials play an important role, thus exhibiting a major dependence of their properties on the number of layers. As $MoTe_2$ shows novel phase change properties and electron-phonon coupling, it is of utmost importance to investigate their layer dependence. Temperature-dependent Raman studies were performed also on exfoliated flakes of pure 1T′-$MoTe_2$ to study the phase transition as a function of flake thickness. The thicknesses of the exfoliated samples were confirmed by optical contrast of the images (see Supplemental material [47] Fig. S7) obtained using an optical microscope fitted with a 100× objective lens, atomic force microscopy (see Figures S8 and S9 in Supplemental material [47]), and also from the Raman spectra obtained at room temperature. Though most of the Raman modes show negligible thickness dependence, a strong dependence is observed for the mode marked $P_{9B}$. As shown in Figure 6(a) (and Figure S10 in Supplemental material [47]), the mode $P_{9B}$ shifts from ~ 270 cm$^{-1}$ for monolayer flakes to ~ 260 cm$^{-1}$ in the bulk. This systematic redshift with increase in flake thickness is consistent with prior experimental and theoretical reports and confirms the layer-thickness in our exfoliated flakes [12,54,55]. Further, monolayer $MoTe_2$ shows a significant broadening of all the modes with an overall change in the lineshapes thus matching well with a previous report [54]. Importantly, with the decrease in thickness of the flakes supported on $SiO_2$/Si substrates, we have observed an increase in the intensity of the $SiO_2$ mode at ~ 303 cm$^{-1}$ (Figure S10 in Supplemental material [47]) thus further corroborating the thickness variation of the flakes.

Figures 6(b) and 6(c) show a comparison between the Raman spectra (phonons near 129 cm$^{-1}$) taken at room temperature (300 K) and at 80 K for the flakes of different thicknesses. The phase transition characterized by the splitting of the mode $P_6$ was prominent in the exfoliated bulk flake with thickness ~ 50 nm and partially evident for the flake of thickness ~ 20 nm. However, the phase transition could not be observed for any of the exfoliated flakes with a thickness ≤ 10 nm including the monolayer, bilayer, and a few-layered flakes as is evident from the absence of the splitting of the $P_6$ mode. Therefore, our Raman measurements verify that below a critical thickness (~ 10 nm), all flakes of MoTe$_2$ show a suppression of the phase transition, stabilizing in the high temperature 1T′ phase in the temperature range from 80 K to 400 K. Here, we verify the proposition of hole doping as the reason for stabilization of the 1T′ phase in thin flakes of MoTe$_2$ by invoking the electron-phonon coupling in the system. Signatures of electron-phonon coupling in MoTe$_2$ were observed in the form of anomalous behavior of the mode $P_1$ as discussed in section 3.3. Figure 6(d) shows the temperature dependence of $P_1$ mode frequency for the exfoliated flakes with various thicknesses. The anomaly in the trend shown by the mode frequency gets suppressed with decreasing flake-thickness in very thin (a few layer) flakes and is finally transformed to the normal behavior for the atomically thin monolayer MoTe$_2$. Therefore, we observe a systematic suppression of the electron-phonon coupling (which we have proposed as the origin of the anomaly) with the decrease in the flake-thickness.

*3.5 Effect of charge doping on the phase transition*

Pawlik *et al*. [54] performed layer-wise ARPES measurements on 1T′-MoTe$_2$ and reported an increasing extent of electron doping with decreasing flake thickness. Kim *et al*. [33], on the other hand, theoretically predicted the stabilization of T$_d$ or 1T′ phases of MoTe$_2$ by electron or hole doping, respectively. Notably, He *et al*. [30] reported that below a critical thickness, MoTe$_2$ remains in the T$_d$ phase even at room temperature which is in contrast to our observations discussed above. It is important to note that the conditions of exfoliation followed by various groups play an important role in tuning the MoTe$_2$ characteristics. The exfoliated flakes for ARPES measurements by Pawlik *et al*. [54] were prepared in nitrogen atmosphere followed by experiments performed under ultrahigh vacuum conditions, while the Raman experiments by He *et al*. [30] were performed on flakes with a capping-layer of h-BN which protected the flakes from exposure to atmosphere. On the other hand, our exfoliated flakes were transferred on to SiO$_2$/Si substrates in normal atmospheric conditions. It is well known that exposure to atmospheric oxygen and moisture substantially hole dopes 2D TMDs [57-63]. We performed experiments on exfoliated flakes which were exposed to the ambient atmosphere to investigate the effect of spontaneous hole doping on the phase transition. Therefore, as predicted in the theoretical work by Kim *et al*. [33], hole doping should lead to 1T′ as the stable phase throughout the temperature, as observed in our Raman data shown in Figure 6. Our observation of 1T′ phase in the temperature range of 80-400 K (i.e. even at temperatures < 250 K) is in sharp contrast to the report of $T_d$ phase by He *et al.* [30] even at high temperatures (> 250 K). Importantly, it is to be noted that our results and the report by He *et al.* [30] are completely in line with the predictions of Kim *et al.* [33]. Though it has been predicted that an intrinsic electron doping is expected in thin flakes of 1T′-MoTe$_2$ [54], this can be very easily overshadowed by a stronger hole doping due to an exposure to atmosphere. This happens because a decrease in the thickness is accompanied by an increase in surface to volume ratio, effectively exposing more surface to the atmosphere and causing stronger effects due to hole doping. There are reports of layer-dependent properties of 1T′-MoTe$_2$ which clearly show the un-split Raman mode near 129 cm$^{-1}$ up to the limit of monolayer, revealing that their flakes always remained in the 1T′ phase irrespective of the number of layers in thinner flakes [64,65]. These reports are also based on samples which were exfoliated under normal atmospheric conditions and without any capping layer on top, thereby resulting in hole doping of their samples.

In order to study the evolution of the electron-phonon coupling, we define a parameter $\delta_{P1}(= \omega_{300\,K} - \omega_{80\,K})$, (See Supplementary Note 7, and Supplementary Figure S20 in Supplemental material [47]) which is the shift in the frequency of the P$_1$ mode at 80 K with respect to its position at room temperature (300 K). As discussed earlier, this shift has contributions from quasi-harmonic and anharmonic terms as well as electron-phonon coupling. However, in absence of electron-phonon coupling, the parameter ($\delta_{P1}$) must show a negative value in line with the anharmonicity theory. Therefore, a positive value for the parameter would represent a dominant contribution from the electron-phonon coupling. It may be explained in a way that an effective increase in the electron population (an effective electron doping) would enhance the electron-phonon coupling further, giving rise to an increase in the parameter $\delta_{P1}$. Figure 7 shows the evolution of the parameter $\delta_{P1}$ for the flakes of different thicknesses. Before attempting to explain the observed trend of $\delta_{P1}$, let us discuss about the different sources of electron and hole doping that are relevant in some detail. As pointed out by Pawlik *et al.* [54] from their ARPES measurements, with decrease in the flake thickness, the Fermi level shows a steady upshift, resulting in n-type (electron) doping. The source of electron doping was attributed to Te-deficiencies and structural defects that could be either induced by the cleavage (exfoliation) process. Again, as discussed earlier, the exposure of the 2D flakes of MoTe$_2$ to atmospheric oxygen and moisture results in hole doping. The mechanism of hole doping in thin flakes of graphene and TMDs have been explained well by Levesque *et al.* [62] and Wang *et al.* [63], respectively. The atmospheric moisture forms a water adlayer (on the 2D flake) in which oxygen molecules get solvated, thereby forming an $O_2/H_2O$ redox couple. This is followed by transfer of electrons from the 2D flake to the $O_2/H_2O$ redox couple, resulting in a hole doping of the 2D flake. Therefore, we may summarize the different possibilities of electron/hole doping as: (i) electron doping ($E_1$) induced by Te-deficiencies and structural defects [54] upon exfoliation, and (ii) hole doping ($H_1$) due to exposure to the atmospheric oxygen and moisture [62,63]. The competition between the terms $E_1$ and $H_1$ effectively electron dopes or hole dopes the flakes of various thicknesses due to the synthesis conditions (See Supplementary Note 8 of Supplemental material [47]). The contribution from the term $H_1$ is confined to the top layer alone (which is exposed to the atmosphere), while the contribution from the term $E_1$ is present in the entire lattice (including all the layers underneath). Therefore, as we move from bulk to thinner layers with exfoliation, in the beginning we expect to see an effective electron doping (and hence an increase in the parameter $\delta_{P1}$) with decreasing thickness, however, for very thin flakes, the term $H_1$ will dominate resulting in an effective hole doping of the monolayer flakes. The dominance of $H_1$ over $E_1$ in very thin layers (below 4-5 layers) is because of the increase in surface area to volume ratio. As a result, we observe a decrease in the parameter $\delta_{P1}$ for thin layers and in case of a bilayer flake it goes to zero and finally becomes negative (which implies a complete suppression of electron-phonon coupling) for the monolayer flake of $MoTe_2$. While the hole doping induced by exposure to atmospheric oxygen and moisture has been clearly observed to stabilize the thin layers of $MoTe_2$ in the 1T′ phase, we also verified that electron doping can stabilize the $T_d$ phase for thin flakes of $MoTe_2$. We achieved electron doping by exposing pure $MoTe_2$ flakes of various thicknesses to $NH_3$ vapor. Exposure to $NH_3$ gas has been predicted theoretically to electron dope 2D layers of TMDs by acting as an electron donor [58]. Figure 8 shows the effect of $NH_3$ induced electron doping on the bulk (~ 50 nm), 5-layer, trilayer, and bilayer flakes of $MoTe_2$. It can be clearly seen that exposure to $NH_3$ does not affect the phase of the bulk flake (~ 50 nm). However, exposing the 2L, 3L, and 5L flakes to $NH_3$ for 10 minutes resulted in a splitting of the $P_6$ mode, therefore, confirming the reappearance of the $T_d$ phase even at room temperature. It may be noted that charge doping in graphene [66] and other TMDs [67] have been reported to show redshifts or blueshift in the frequencies of certain phonons. Therefore, inequivalent charge doping of various layers in a few-layered flake may possibly result in variable shifts in phonon frequencies contributed by different layers, giving rise to a splitting in the mode. However, in our results, inequivalent doping of various layers as an origin of the splitting of the $P_6$ mode can be completely ruled out due to the following reasons: (i) we do not observe any splitting of the $P_6$ mode for bulk flakes (> 50 nm), and (ii) there is no measurable change observed in any other modes due to exposure to $NH_3$ vapor (See Supplementary Note 6 of Supplemental material [47]). Therefore, we may conclude that the electron doping induced by exposure to $NH_3$ vapor results in the stabilization of the $T_d$ phase down to atomically thin flake thicknesses.

### 3.6 Electron doping of MoTe$_2$ by incorporating Re

In order to study the effect of electron doping on the phase transition of MoTe$_2$, we have performed Raman studies on exfoliated thin flakes of Mo$_{1-x}$Re$_x$Te$_2$; x=0.1 single crystal. Figure 9(a) shows a comparison between the Raman spectra (near 129 cm$^{-1}$) obtained at room temperature and at 80 K. In contrast to the exfoliated thin flakes of pure MoTe$_2$, where phase transition is quenched in flakes of thickness ≤ 10 nm, we find that the 10% Re doped flakes of any given thickness, even down to monolayer show a clear splitting of the 129 cm$^{-1}$ mode at low temperatures indicating the existence of the phase transition from 1T′ to the $T_d$ phase. This is expected as the Re doping at the Mo sites has been reported to show electron doping in MoTe$_2$ [48]. As the exfoliation of the Mo$_{1-x}$Re$_x$Te$_2$; x=0.1 single crystal was performed under similar atmospheric conditions as the pure flakes, we might expect some amount of hole doping to be induced due to the exposure to atmospheric oxygen and moisture. In addition to the sources of electron/hole doping that were present for the pure flakes of MoTe$_2$, here we observe an additional contribution as a result of Re doping. This results in a competition between the following sources of electron/hole doping: (i) electron doping ($E_1$) induced by Te deficiencies and structural defects [54] upon exfoliation, (ii) electron doping ($E_2$) due to the incorporation of Re at the Mo sites of the lattice [48], and (iii) hole doping ($H_1$) due to exposure to the atmospheric oxygen and moisture [62,63]. Unlike flakes of pure MoTe$_2$ which were influenced by the terms $E_1$ and $H_1$ only, the additional contribution $E_2$ leads to an effective electron doping in Mo$_{1-x}$Re$_x$Te$_2$; x=0.1 thin flakes. A qualitative comparison between the competing terms and the effective electron doping is discussed in Supplementary Note 8 of Supplemental material [47]. We can further verify the effective electron doping from the evolution of the electron-phonon coupling in the flakes of Mo$_{1-x}$Re$_x$Te$_2$; x=0.1 with thickness. As discussed earlier, a positive value of the parameter $\delta_{P1}$ (which is a measure of the anomaly in the mode $P_1$) may be considered as a result of effective electron doping. In Figure 9(b), we observe that for all thicknesses of exfoliated $Mo_{1-x}Re_xTe_2$; x=0.1 flakes, $\delta_{P1}$ is always positive which may be interpreted as an effective electron doping in all the flakes down to atomically thin layers (See Supplementary Note 7, and Supplementary Figure S21 in Supplemental material [47]). We also observe an initial increase in $\delta_{P1}$ with thinning down of flakes down to a critical thickness of ~ 7-8 layers. This is expected because for bulk flakes, the contributions $E_1$ and $E_2$ (which are present in the entire lattice) dominate over the contribution of $H_1$ (which affects only the top layer that is exposed to air). However, in case of thinner flakes (< 7 layers) which constitute a few layers only, the relative contribution from $H_1$ would increase as a function of decreasing thickness due to an increase in the surface to volume ratio, thus resulting in an effective decrease in $\delta_{P1}$. Notably, even in these thinner flakes the net contribution due to $E_1$ and $E_2$ still dominates over that of the $H_1$ and, hence, the $\delta_{P1}$ remains positive indicating the presence of electron-phonon coupling irrespective of the thickness of the flake down to monolayer.

## 4. Discussion

In summary, we have performed extensive Raman measurements on 1T′-MoTe$_2$ single crystals of different thicknesses and studied various aspects which may be categorized into the following main points:

Firstly, the bulk flakes of pure MoTe$_2$ reveal a hysteretic phase transition at ~ 250 K, which we confirmed using charge transport and Raman measurements. The phase transition is accompanied by a complete renormalization of the phonons in the system due to the breaking of inversion symmetry. The detection of the inversion symmetry breaking is very important as it ensures that the low temperature phase is a Weyl semimetal. Though the confirmation of the Weyl semimetallic phase may come from ARPES measurements, the reports on ARPES data are inadequate for the confirmation. Weyl semimetallic phase is characterized by the presence of Fermi arcs connecting the Weyl nodes having opposite chiralities [19], signatures of which were reportedly seen in the ARPES measurements on $T_d$-MoTe$_2$ [24]. However, thermal broadening makes it difficult to prove the absence of the Fermi arcs in the 1T′ phase. Therefore, a more prominent proof of inversion symmetry breaking is desired, which is why the Raman detection of the phase transition is important.

Secondly, it was possible to tune the phase transition temperature in bulk MoTe$_2$ by incorporating Re doping (10%) at the Mo sites of MoTe$_2$ by ~ 30 K.

Thirdly, our studies revealed the suppression of the phase transition in thin flakes of thickness ≤ 10 nm. However, it was possible to attain the desired phase (1T′ or T$_d$) at any particular temperature (between 80 to 400 K) and for any arbitrary thickness by charge doping. In order to verify this prediction by Kim *et al.* [33], the exfoliated flakes were hole doped externally by exposure to ambient air. On the other hand, the electron doping was achieved by two methods: exposure to NH$_3$ vapor (external doping), and doping with Re at the Mo sites of the MoTe$_2$ lattice (internal doping). Both the techniques produced similar effects, showing a stabilization of the T$_d$ phase. We, therefore, propose that tunable charge doping can be an important way of stabilizing either of the topological phases for flakes of any random thickness, as has been recently observed in WTe$_2$ [68]. It is also important to note that while the first technique of external charge doping by exposure to atmosphere or ammonia vapor is well understood, the internal doping by incorporation of Re at the Mo sites is a complex situation. This is because the Re doping into the crystal may also introduce local distortions in the crystal structure, giving rise to new properties and phenomena. For example, as observed by Mandal *et al.* [48], the Re doping results in internal pressure in the crystal which may be responsible for the elevation of the superconducting transition temperature. Similarly, local strain induced by incorporation of Re in the crystal gives rise to a pseudo magnetic field in MoTe$_2$ [49]. Hence, the complete understanding of the phase transition pathways in the internally doped samples demands further theoretical investigation.

Finally, we have observed clear signatures of electron-phonon coupling in MoTe$_2$. MoTe$_2$ is a low-T$_c$ superconductor, which shows an increase in T$_c$ on doping with Re at the Mo sites [48]. Superconductivity is often related to anomalies in phonon dispersions, which are manifested by anomalous temperature dependence in the frequencies and linewidths of the concerned phonons as has been reported for carbon nanotubes [41] and YNi$_2$B$_2$C [42]. The anomalous temperature behavior has been established by various theoretical and experimental (neutron scattering experiments) reports to be a signature of electron-phonon coupling [41-46]. We have discussed in detail about the evolution of electron-phonon coupling as a function of flake thickness taking into account the effects of electron/hole doping.

## 5. Conclusion

Our work shows a complete picture of the phonon renormalization that takes place as a result of the structural phase transition in MoTe$_2$ arising from an inversion symmetry breaking. We believe that our study of phase transition phenomena as a function of layer thickness (taking into account the effects of charge doping) will be vital to understanding the relevance and limitations of using MoTe$_2$ in various applications. Further, we verified the switching between the two topologically important phases of MoTe$_2$ by simple techniques of internal and external charge doping. We believe that our Raman-based study of the phase transition as well as of electron-phonon coupling and the possibility of tuning these in MoTe$_2$ will motivate further studies, both theoretical and experimental, exploring the physical properties and topological phenomena of MoTe$_2$.

**Additional notes**

While our work was under review, there have been two recent reports relevant to our work as follows:

1. Xiao *et al.* [68] have demonstrated an electrically driven stacking order transition from the non-centrosymmetric T$_d$ phase to the centrosymmetric 1T′ phase of WTe$_2$, revealing a similar trend of stabilizing the latter with hole doping, and vice versa, as observed in our work on MoTe$_2$. The report also claims that such an electrically-driven stacking order transition may be employed to realise Berry curvature memory.
2. Zhang *et al.* [69] demonstrated signatures of electron-phonon coupling in bulk 1T′-MoTe$_2$ and related it to the topological phase transition. Our work brings out further novelties with a systematic study of the phase transition and electron-phonon coupling in 1T′-MoTe$_2$ as a function of varying thickness and charge doping.

**Acknowledgements**

Funding from DST/SERB (Grant No. ECR/2016/001376, Grant No. YSS/2015/001799, and Grant No. CRG/2019/002668), Nanomission (Grant No. SR/NM/NS-84/2016(C)) and DST-FIST (Grant No. SR/FST/PSI-195/2014(C)) are acknowledged.

**Author contributions**

S.S. and R.P.S. conceived the idea. S.P., S.K., A.R. performed mechanical exfoliation of flakes and Raman measurements. S.P., S.K. analysed the data. M.M., S.M. with guidence from R.P.S. synthesised the single crystal flakes. S.S. supervised the research work. S.P., S.K., S.S. jointly prepared the manuscript in consultation with all the authors.

**Figures:**

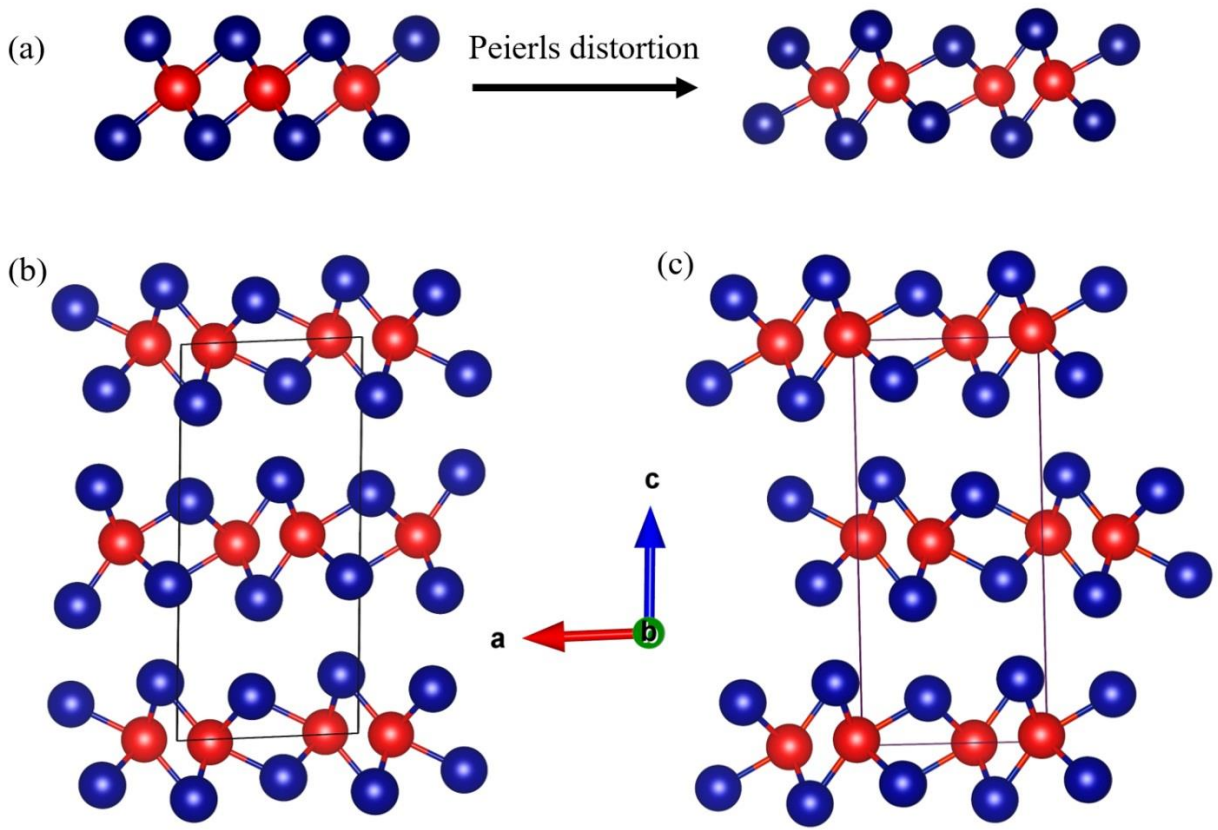

Fig. 1. (a) Peierls distortion of $MX_2$ from 1T to 1T′ phase; (b) Crystal structure of 1T′ and (c) $T_d$ phases of $MoTe_2$.

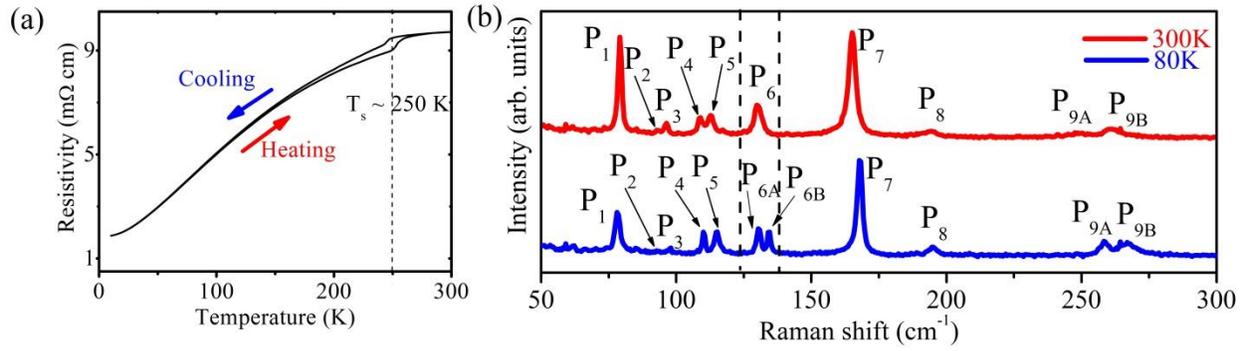

Fig. 2. (a) Resistivity vs temperature for bulk 1T′-MoTe$_2$ showing hysteresis near 250 K; (b) Raman spectra of 1T′-MoTe$_2$ taken at 300 K and 80 K revealing the splitting of the P$_6$ mode to P$_{6A}$ and P$_{6B}$ at low temperature due to breaking of inversion symmetry.

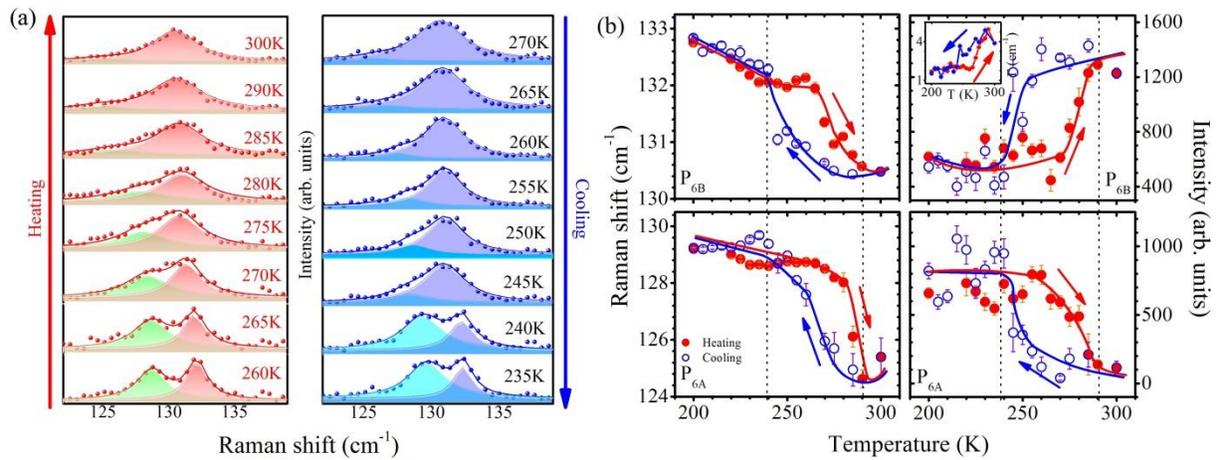

Fig. 3. (a) Raman spectra showing the evolution of $P_6$ mode in the heating and cooling cycles in response to the $1T'\leftrightarrow T_d$ phase transition; (b) Hysteresis in the frequencies and integrated intensities of modes $P_{6A}$ and $P_{6B}$. Inset shows the hysteresis in linewidth of mode $P_{6B}$. The red and blue data points represent the heating (red solid circles) and cooling (blue open circles) cycles. The solid connecting lines are guides to eye for clearly visualizing the hysteresis.

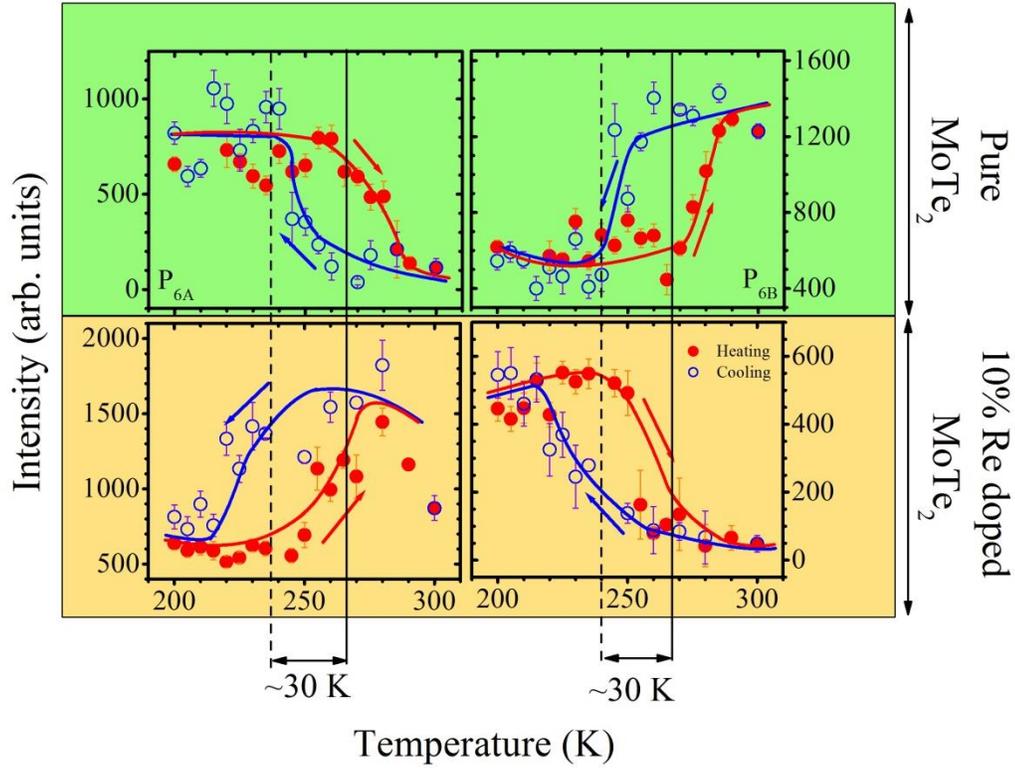

Fig. 4. The hysteresis in intensities of modes $P_{6A}$ and $P_{6B}$ in pure $MoTe_2$ and $Mo_{0.8}Re_{0.2}Te_2$ crystals. The red and blue data points represent the heating and cooling cycles. The solid (dashed) vertical line shows the centre of the hysteresis loop for the pure (10% Re doped) $MoTe_2$ crystals respectively. We observe a shift of ~ 30 K due to Re doping.

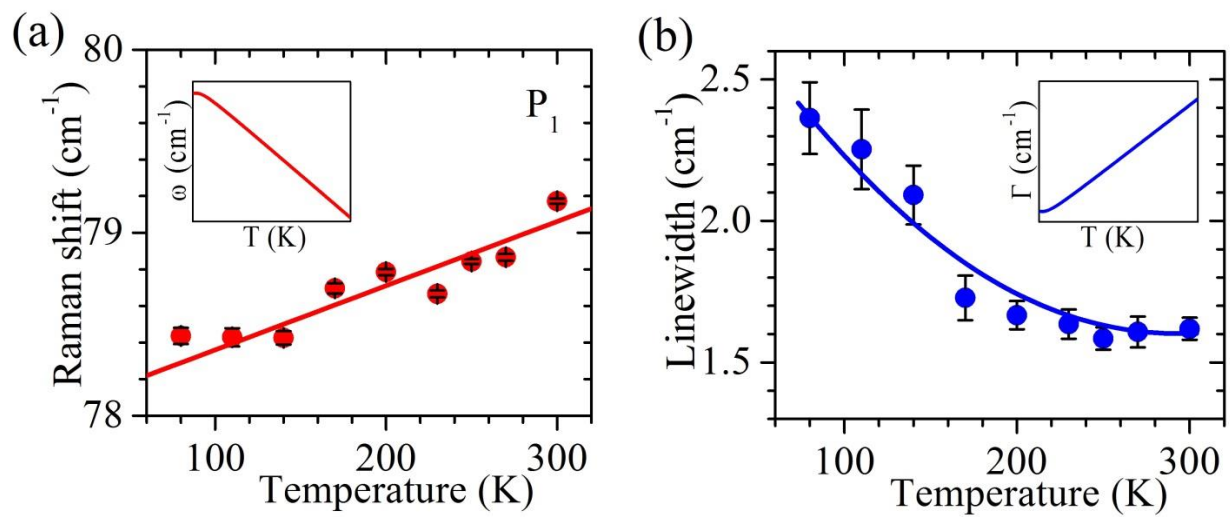

Fig. 5. Anomalous behavior of (a) frequency and (b) linewidth of the $P_1$ mode as a function of temperature. The insets show the typical behaviors expected as per phonon anharmonicity represented by equations (1) and (2) in the text.

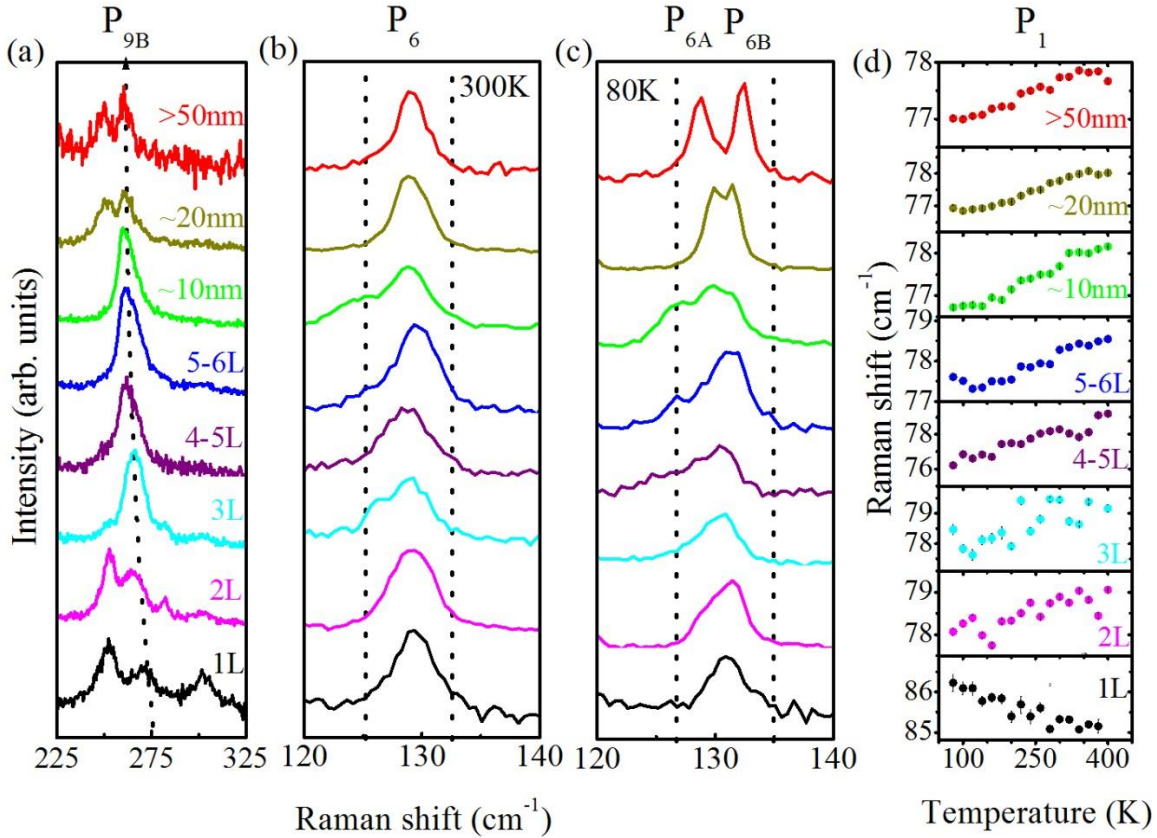

Fig. 6. (a) Dependence of the Raman mode P$_{9B}$ on the layer thickness of 1T′-MoTe$_2$; (b,c) Comparison of the Raman spectra (near ~ 129 cm$^{-1}$) at 300 K and 80 K, showing suppression of phase transition for all exfoliated flakes < 10 nm. The phase transition characterized by the splitting of mode P$_6$ appears in the flakes with thicknesses ~ 20 nm and ~ 50 nm; (d) Anomalous behavior of frequency as a function of temperature showing suppression of electron-phonon coupling with the decrease in flake thickness. The error bars are within or comparable to the size of the data points.

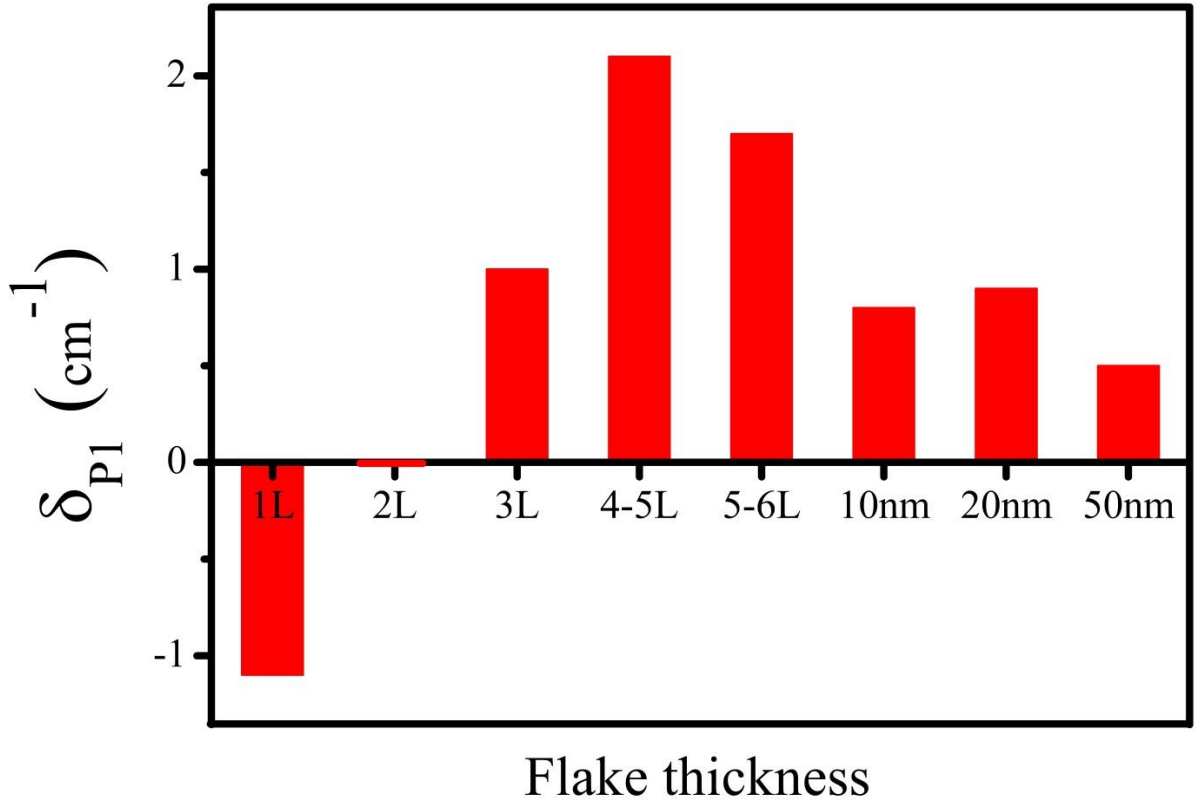

Fig. 7. The evolution of the parameter $\delta_{P1}$ for pure $MoTe_2$ flakes of different thicknesses. The parameter $\delta_{P1}$ is defined as the shift of the mode $P_1$ at 80 K with respect to its position at room temperature (300 K). A positive value of the parameter signifies a dominant electron-phonon coupling. We observe an initial increase in the parameter with decreasing flake thickness, while for very thin layers the parameter shows decreasing trend, with reversal from positive to negative sign for monolayer $MoTe_2$ implying the absence of the coupling in monolayer.

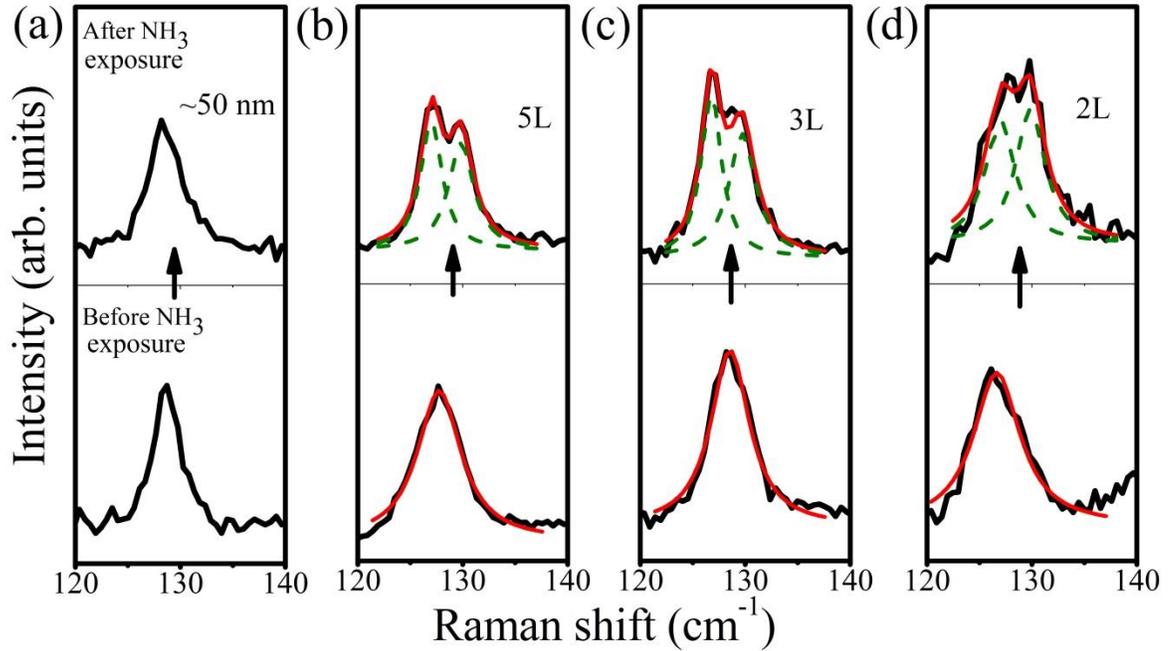

Fig. 8. Effect of exposure to NH$_3$ vapor for 10 minutes at room temperature for (a) bulk (~ 50 nm), (b) 5-layered, (c) 3-layered, and (d) 2-layered flakes. The bulk flake shows no change and remains in the 1T′ phase, while the thin-layered flakes clearly show two modes near 129 cm$^{-1}$, which indicates the stabilization of the T$_d$ phase at room temperature. The spectra for 5L, 3L, and 2L flakes are fitted with Lorentzian multi-functions (green dashed lines) to clearly show the splitting that appeared after exposure to NH$_3$ vapor.

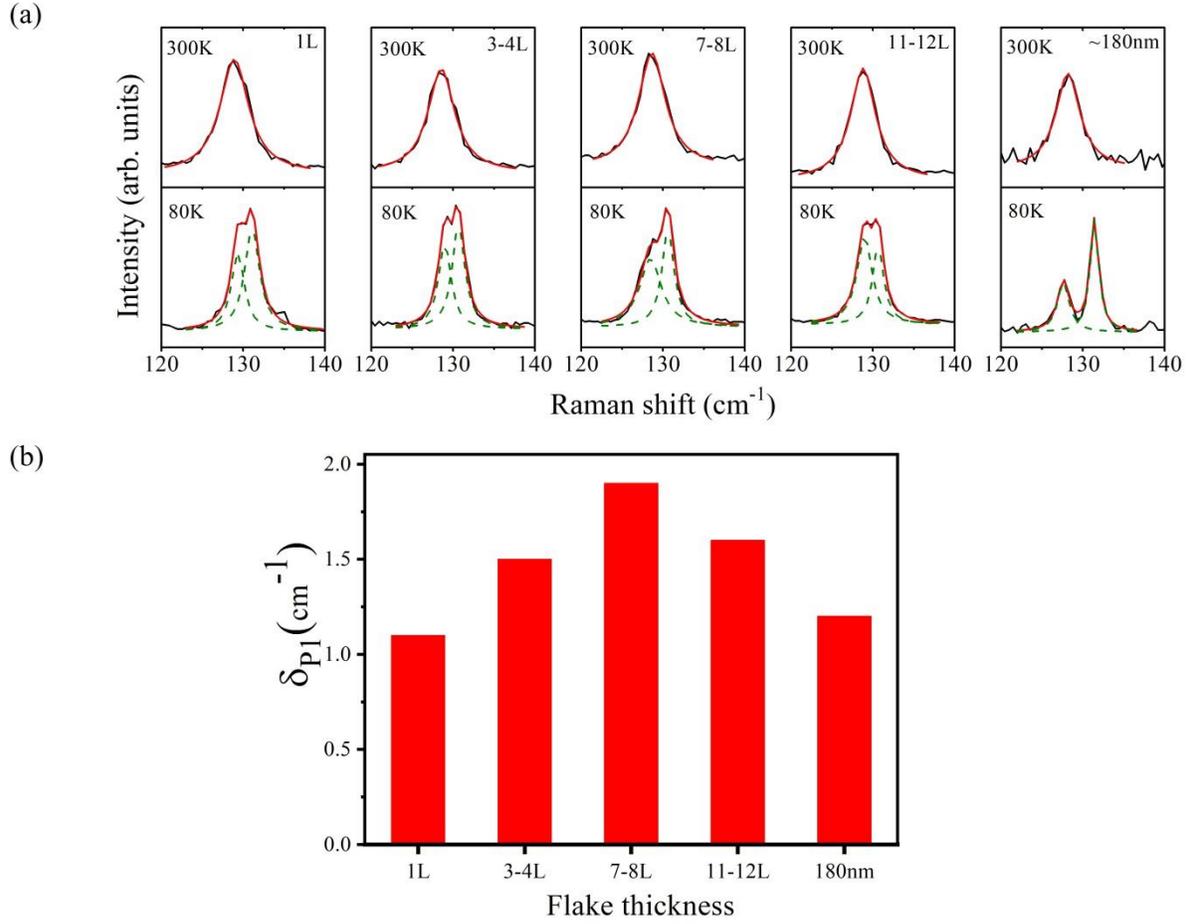

Fig. 9. Effect of electron-doping induced by doping Re at Mo sites of MoTe$_2$. (a) The comparison of Raman spectra obtained at 300 K and at 80 K clearly shows a splitting of the P$_6$ mode at low temperature, indicating the phase transition from 1T′ to T$_d$ phase for flakes of any random thickness. The spectra are fitted with Lorentzian multi-functions (green dashed lines) to clearly show the splitting that appeared at 80 K, (b) The parameter δ$_{P1}$ for 10% Re doped MoTe$_2$ flakes of various thicknesses. The parameter follows a similar trend as in pure crystal, but, due to the increase of electron population induced by Re doping, it never becomes negative, down to atomically thin layers thus implying the presence of electron-phonon coupling in flakes of any given thickness.

# Tailoring the Phase Transition and Electron-Phonon Coupling in 1T′-MoTe$_2$ by Charge Doping: A Raman Study

Suvodeep Paul, Saheb Karak, Manasi Mandal, Ankita Ram, Sourav Marik, R. P. Singh, and Surajit Saha[*]

([*]surajit@iiserb.ac.in)

*Department of Physics, Indian Institute of Science Education and Research Bhopal, Bhopal, 462066, India*

This Supplementary information contains additional data including single crystal X-ray diffraction and Raman data of bulk single crystal of pure and 10% Re doped 1T′-MoTe$_2$. There is a brief discussion on the temperature dependence of the different Raman modes. The hysteresis in linewidth of mode P$_{6A}$ in pure MoTe$_2$ is shown. The temperature stacks of Raman data revealing the phase transition in Re doped MoTe$_2$ are shown. The typical temperature behaviour of mode frequencies and linewidths predicted from anharmonic approximation is also discussed. Atomic Force Microscopy data and optical images obtained using objective microscope help in determining the thickness of exfoliated flakes. Additionally, there are discussions on the measurement of the parameter δ$_{P1}$, which has been defined to estimate and qualitatively compare the electron-phonon interactions in different flakes of pure and Re doped MoTe$_2$.

## Supplementary Note 1:

*Preparation and characterization of single crystal 1T'-MoTe$_2$*

Single crystals of 1T′-MoTe$_2$ (and Mo$_{1-x}$Re$_x$Te$_2$; x=0.2) were prepared by Chemical VaporTransport method as discussed in the main text. Following the preparation, the pure MoTe$_2$ crystals were characterized using PANalytical diffractometer equipped with Cu-K$_\alpha$ radiation. The XRD pattern obtained is shown in Figure S1. The XRD pattern matches well with previous report [1]. The sharp peaks indicate the high degree of crystalline quality of the flakes prepared. The presence of the (001) planes in the XRD indicate that the exposed surface was perpendicular to the c-axis. The inset of Figure S1 shows a photograph of the as-grown 1T′-MoTe$_2$ flake on which experiments were performed. The characterization of the Re doped crystals were reported by Manasi et al. [2] and are not discussed here.

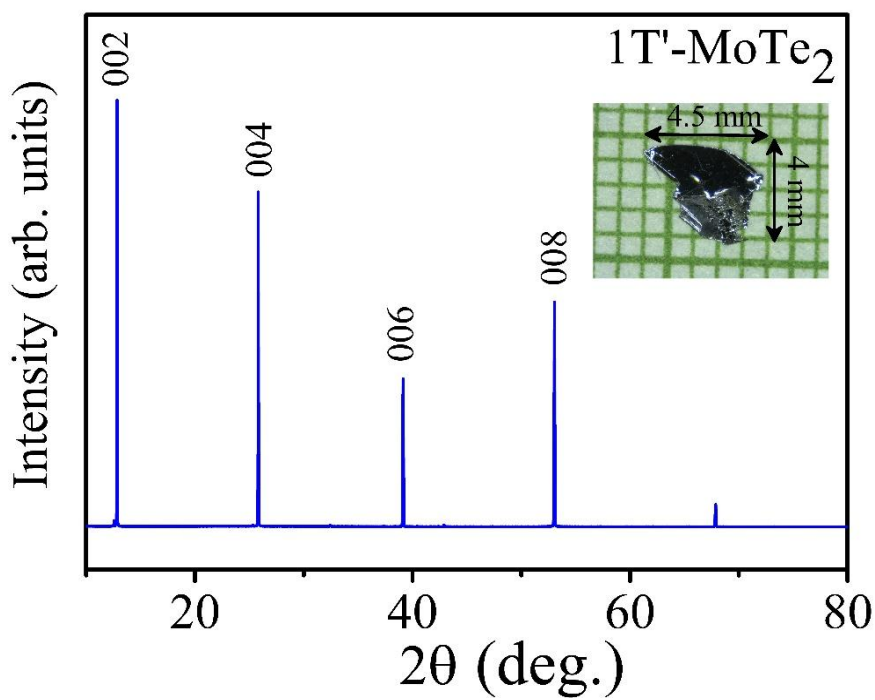

**Fig. S1. The XRD pattern for as-grown single crystal of 1T′-MoTe$_2$. Inset: Photograph of the single crystal 1T′-MoTe$_2$ on which experiments were performed.**

**Supplementary Note 2:**

*Room temperature Raman spectrum of 1T'-MoTe$_2$*

The room temperature Raman spectrum of 1T′-MoTe$_2$ is shown in Figure 2 in the main text. The modes in the spectrum are labelled as P$_1$ to P$_9$. The following table enlists the different modes (P$_1$ to P$_9$) with their corresponding symmetry assignments and frequencies.

| Mode label | Symmetry assignment | Frequency (cm$^{-1}$) |
|---|---|---|
| P$_1$ | $A_g$[3,4,5] | 79 |
| P$_2$ | $B_g$[4] | 93 |
| P$_3$ | $B_g$[4] | 96 |
| P$_4$ | $A_g$[4] | 109 |
| P$_5$ | $A_g$[3,4] | 112 |
| P$_6$ | $A_g$[4] | 129 |
| P$_7$ | $B_g$[3] / $A_g$[4,5] | 165 |
| P$_8$ | $B_g$[4,5] | 192 |
| P$_{9A}$ | $A_g$[3] | 250 |
| P$_{9B}$ | $A_g$[3,4,5] | 260 |

## Supplementary Note 3:

*Hysteretic structural phase transition*

The bulk flakes of 1T′-MoTe$_2$ undergo a first order structural phase transition near 250 K to the low temperature T$_d$ phase. This structural phase transition is reflected in the electrical resistivity and phonon behaviour probed by Raman spectroscopy with a hysteresis. The hysteresis in resistivity is shown in Figure 1(b) and discussed in main text. The phase transition is accompanied by the breaking of inversion symmetry which results in the renormalization of phonons. This gives rise to a hysteretic behaviour in the phonon frequencies, linewidths and intensities of the modes P$_{6A}$ and P$_{6B}$ as is shown in Figure 2(b) and discussed in main text. The hysteresis observed in the linewidth of the mode P$_{6A}$ is poor as compared to the hysteresis in the other parameters. Figure S2 shows the heating and cooling cycles for the linewidth of mode P$_{6A}$ showing a very weak hysteresis.

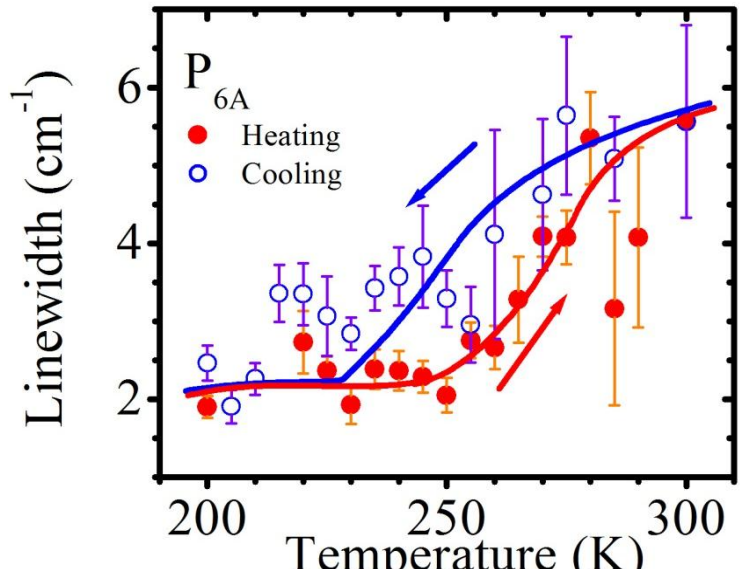

**Fig. S2. Heating and cooling plots for linewidth of P$_{6A}$ mode as a function of temperature.**

The phase transition in Mo$_{1-x}$Re$_x$Te$_2$ (x=0.1,0.3) are also characterized by the appearance of two modes (P$_{6A}$ and P$_{6B}$) near 129 cm$^{-1}$ in the low temperature phase. However, the behaviour of P$_{6A}$ and P$_{6B}$ appear to have reversed (Figures S3 and S4) as we find that the mode P$_{6B}$ appears to vanish in the high temperature phase. This unusual behaviour is not understood well. It is likely that the electron doping (induced by Re doping at Mo sites) of MoTe$_2$ causes a change in the phonon dispersion which leads to reversal of identities of the modes.

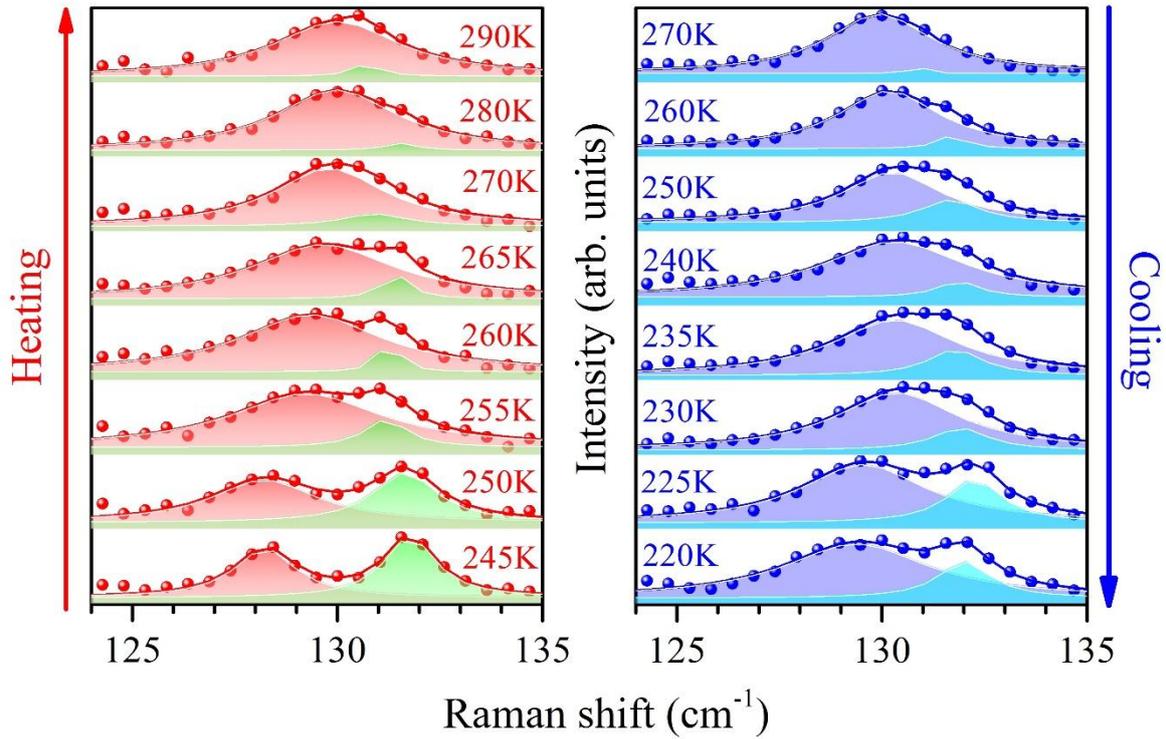

**Fig. S3.** The heating and cooling cycles for the Raman data of $Mo_{0.8}Re_{0.2}Te_2$ near 129 cm$^{-1}$. The phase transition takes place at different temperatures in the heating and cooling cycles, giving rise to a hysteresis.

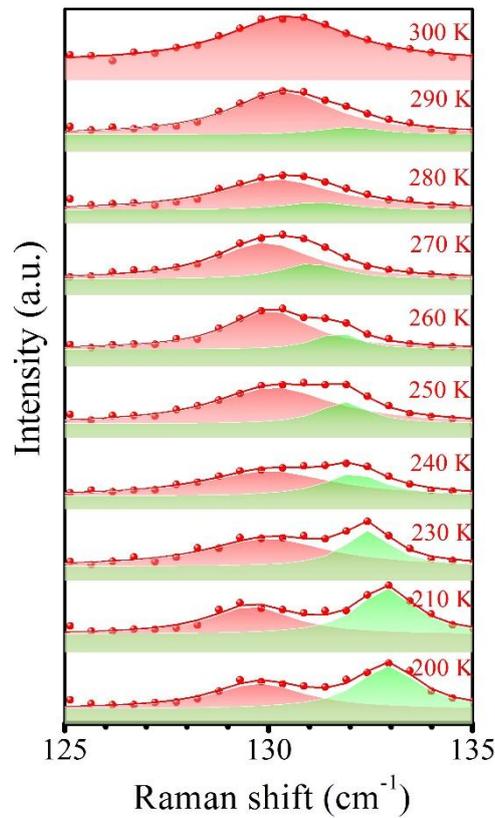

**Fig. S4.** The phase transition in $Mo_{0.7}Re_{0.3}Te_2$ from $T_d$ to $1T'$ phase revealed from Raman data near 129 cm$^{-1}$.

## Supplementary Note 4:

*Expected trends of temperature-dependent Raman data*

As discussed in the main text, the behaviour of Raman modes can be predicted well by considering the anharmonic approximation, which include effects of thermal expansion and inclusion of anharmonic (cubic and quadratic terms) in the Hamiltonian. The self-energy associated with a Γ point LO or TO phonon which decays into acoustic phonons of opposite momenta, has a real part (mode frequency) and an imaginary part (linewidth), both of which are temperature-dependent. The coupling of the zone center optical phonons to the acoustic phonon into which the former decays were studied using various models depending on the various possible decay channels [5]. But, general trend for the mode frequency and linewidth are well explained by the anharmonicity equations (1) and (2) given in the main text. The equations are plotted in insets of Figure 4 of main text. The mode $P_1$ clearly shows an anomaly in both the frequency and linewidth trends as a function of temperature which cannot be explained using the anharmonic approximation and has been attributed to electron-phonon coupling. However, the other modes (modes $P_2$-$P_9$) show the expected trends in both mode frequency and linewidth as shown in Figures S5 and S6.

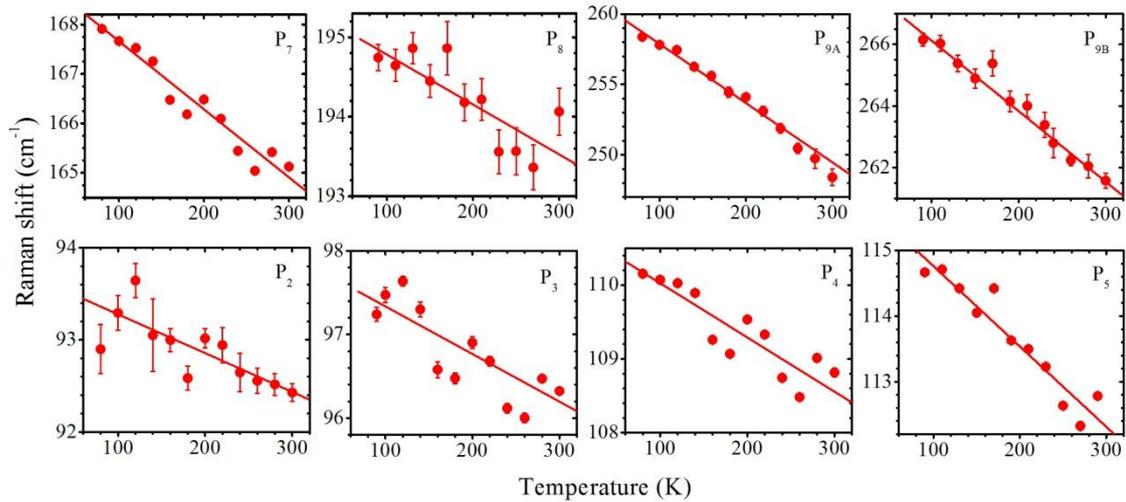

**Fig. S5. Mode frequencies for the modes $P_2$-$P_9$ as a function of temperature. The behavior can be explained by phonon anharmonicity as discussed in the main text. Solid lines are fittings by equation (1) of main text.**

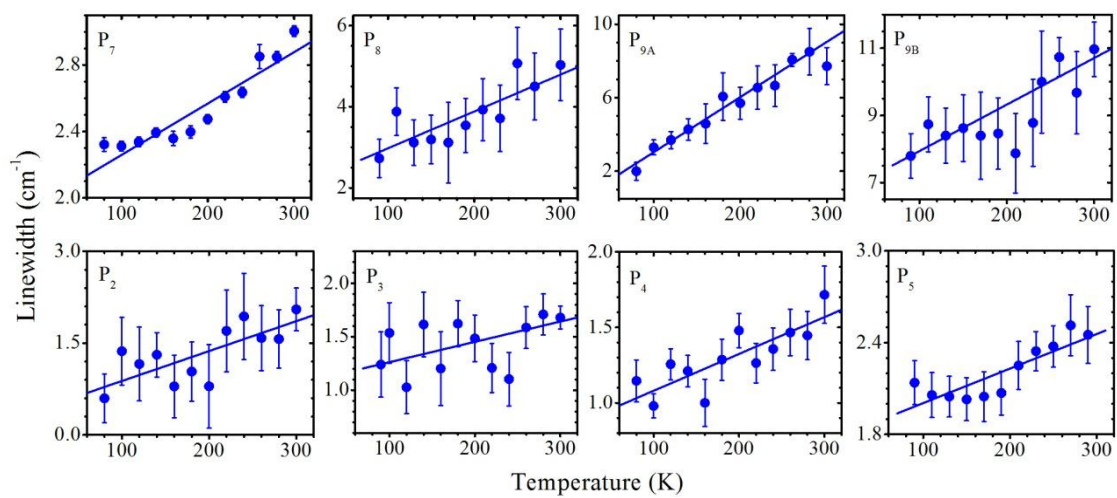

**Fig. S6.** Linewidths of the modes $P_2$-$P_9$ as a function of temperature. Solid lines are fittings by equation (2) of main text, thus suggesting anharmonic behavior of the phonon modes.

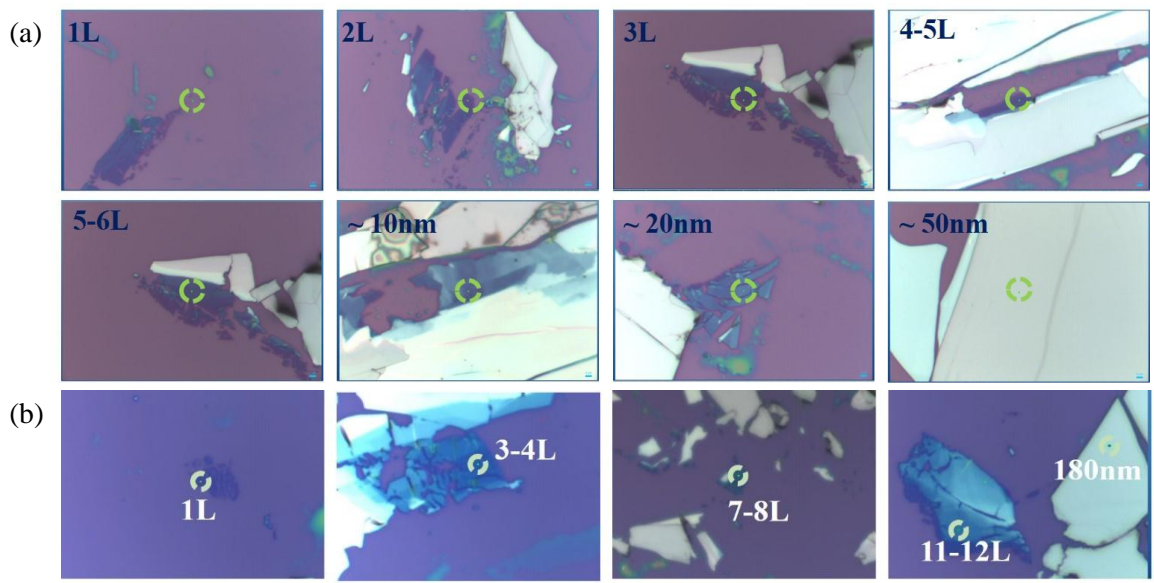

**Fig. S7.** Optical microscopic images of exfoliated flakes of different thicknesses (encircled with green) obtained by micromechanical exfoliation.

## Supplementary Note 5:

*Exfoliated samples and thickness-dependence*

To study the thickness dependence of the phase transition and to observe the effects of atmospheric oxygen and ammonia vapour induced hole and electron doping respectively, we performed micromechanical exfoliation of pure flakes of MoTe$_2$ using scotch tape. Effect of electron doping was also observed by studying thin layered flakes of Re doped MoTe$_2$ crystal, which were again obtained similarly. The obtained layers were transferred on to silicon substrates coated with 300 nm thickness of SiO$_2$ coating. To increase the yield of different layers by the technique, we performed piranha treatment on the substrates before transferring the layers. The thicknesses of the few-layered samples could be confirmed using optical microscopy images which were obtained using a 100× objective lens. The optical images of exfoliated pure and Re doped flakes are shown in Figure S7(a) and S7(b) respectively. The typical sizes of the different layers obtained were of the order of ~10-50 microns. As we were interested to study the critical thickness for suppression of phase transition, we also performed experiments on some bulk exfoliated samples of various thicknesses, which were determined using AFM measurements as shown in Figure S8 and S9 for pure and Re doped flakes. The suppression of phase transition is reflected in the Raman data by the appearance of a single mode (instead of two) at all temperatures in the range 80 K to 400 K. This is clearly shown in Figure 6(b,c) of main text. Figure S10 shows a comparison between the complete Raman spectra of exfoliated layers of 1T′-MoTe$_2$, clearly showing all the modes. We observe no appreciable change in the other modes either again suggesting the suppression of phase transition.

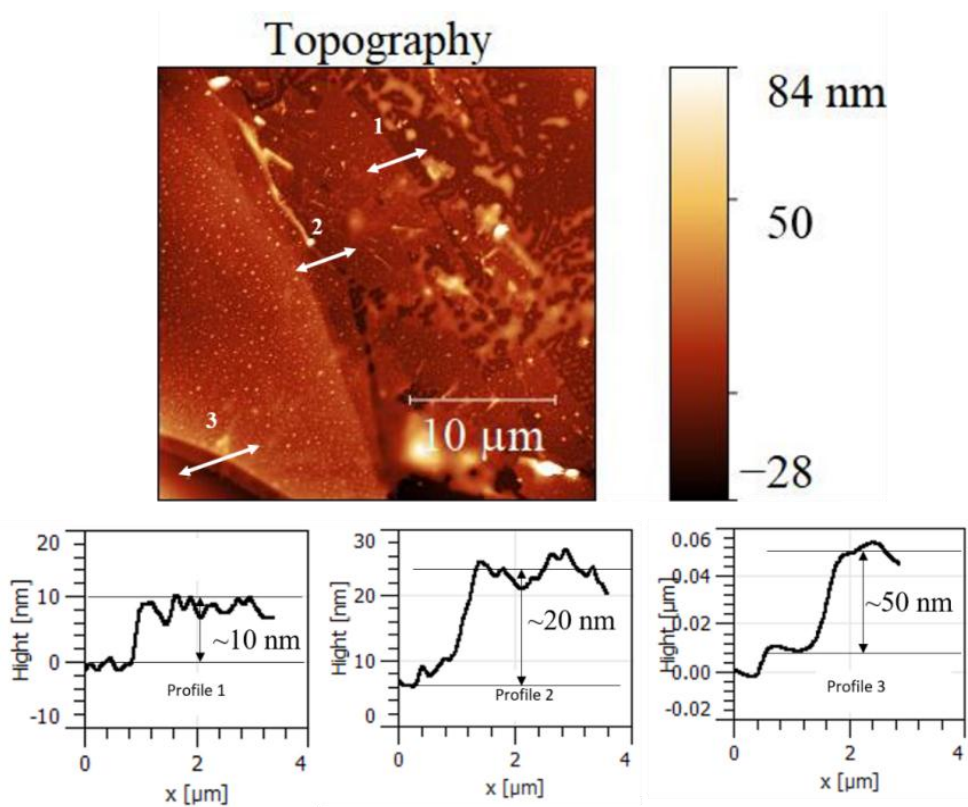

**Fig. S8. AFM images and height profile of few exfoliated flakes of pure MoTe$_2$.**

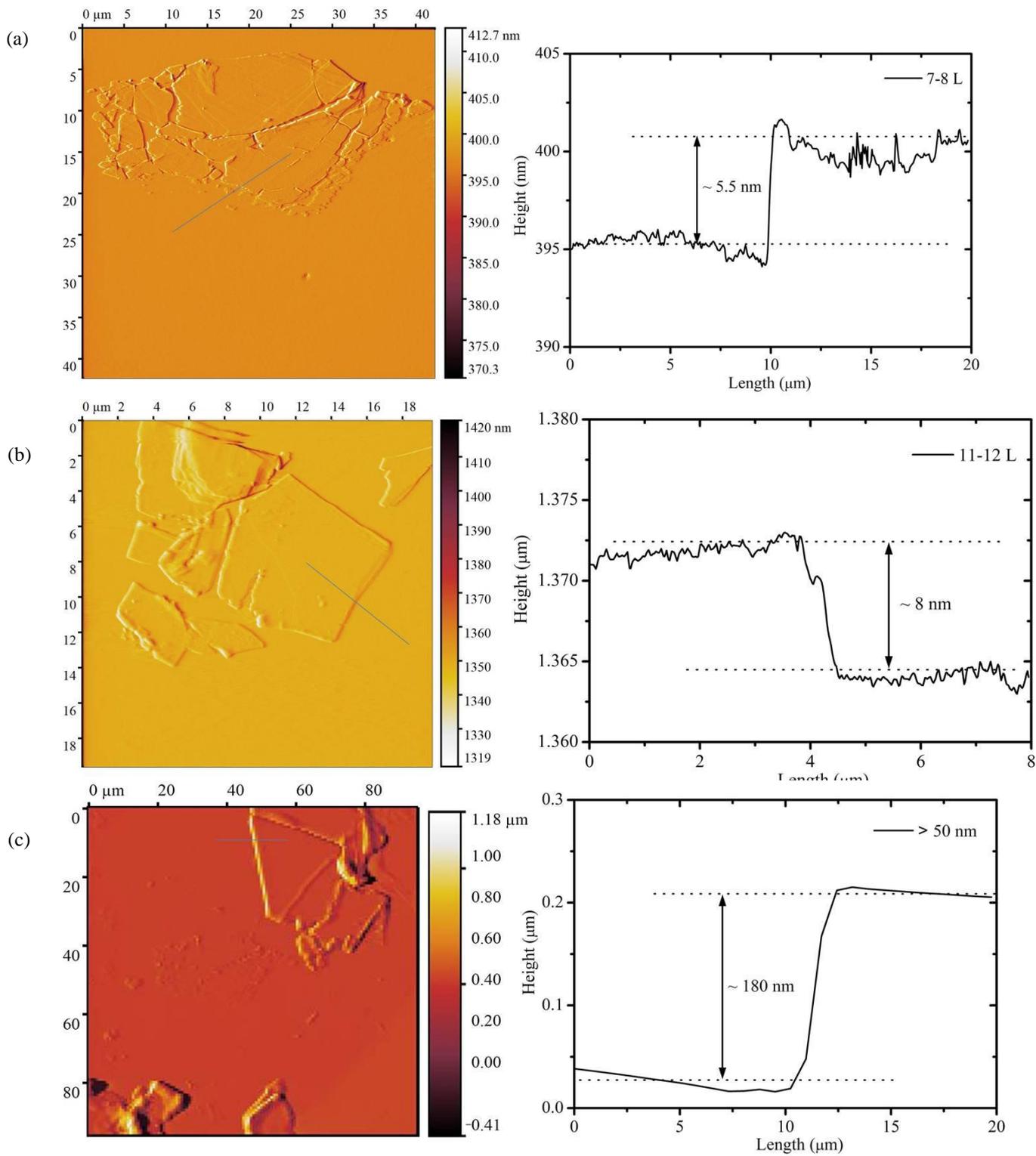

**Fig. S9.** AFM image and corresponding height profiles of exfoliated flakes of 10% Re doped MoTe$_2$. (a) 7-8 layers (5.5 nm), (b) 11-12 layers (8 nm), and (c) 180 nm thick flake.

S9

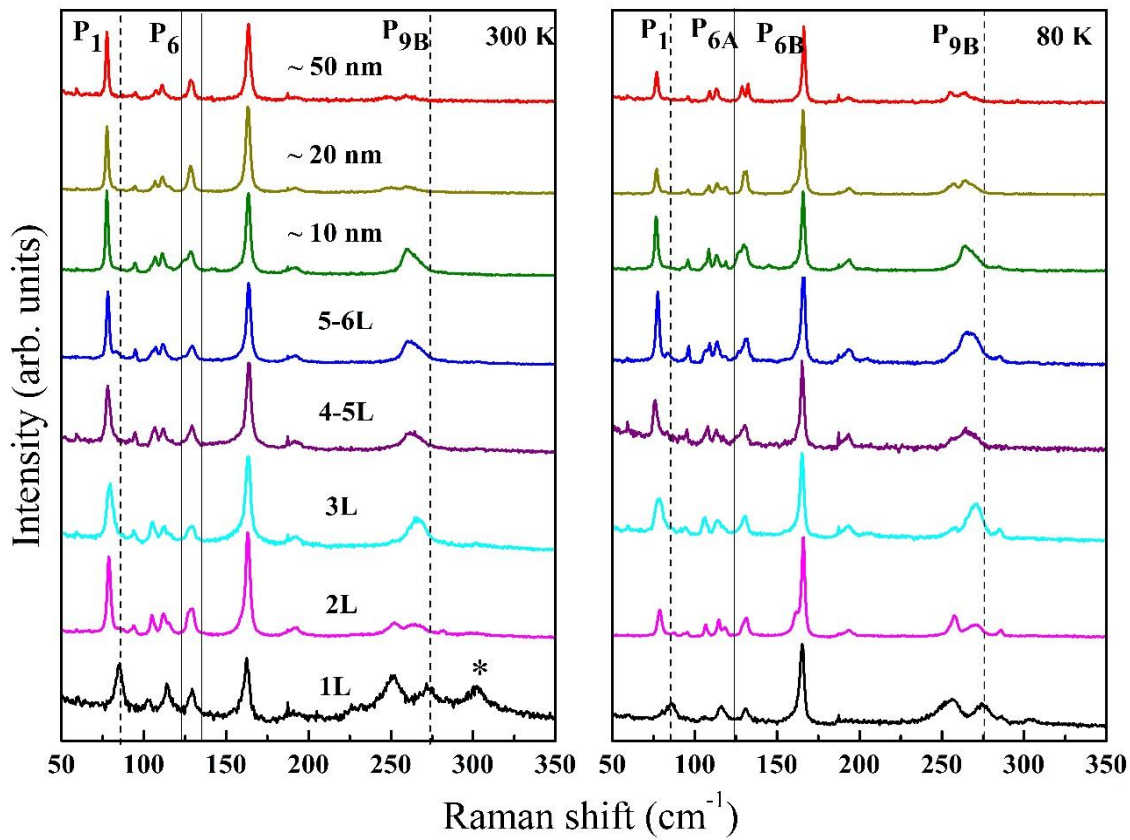

**Fig. S10.** Comparison of Raman spectra for different exfoliated samples at 300 K and at 80 K. The mode $P_{9B}$ shows strong dependence on the flake thickness. The mode labelled * at ~ 303 cm$^{-1}$ for 1L MoTe$_2$ in the left panel (300 K) represents the SiO$_2$ mode from the substrate, and its intensity dimisihes with increase in flake thickbness of MoTe$_2$.

**Supplementary Note 6:**

*Electron and hole doping*

We have experimentally verified that hole (electron) doping can stabilize the QSH 1T′ phase (WSM T$_d$ phase) at room temperature for MoTe$_2$ flakes of any thickness. We achieved hole doping by exposure to atmospheric oxygen and moisture, which has been reported to hole dope TMDs. Again, based on a theoretical report of ammonia being an electron donor, we exposed the exfoliated MoTe$_2$ flakes in ammonia vapour. Though this technique was very effective in doping the MoTe$_2$ flakes, we have observed that over-exposure to ammonia vapour degrades the layers of MoTe$_2$. Therefore, it is not preferable to expose the material to ammonia vapour for a very long time. Thus, we exposed the different exfoliated flakes for 10 minutes only.

Figures S11, S12, S13, and S14 show the effect of electron doping (induced by exposure to NH$_3$ vapour) in bulk, 5-layered, 3-layered, and 2-layered samples respectively. We clearly observe that the bulk flake shows no substantial change on exposure to NH3, while the thin flakes show changes near 129 cm$^{-1}$ regions of the respective spectra. It has been reported that charge doping (induced by environmental effects) are manifested in graphene and different TMDs by the shift in position of phonon modes [7]. An obvious question that may arise

is the possibility that doping of the top surface (due to $NH_3$ vapour exposure) has caused a shift in the mode near 129 cm$^{-1}$, while the layers underneath (which are protected from doping by the top surface) show no such shift, resulting in the appearance of a splitting of the mode. But this is not the case in our data as we find that the bulk flake (having thickness ~ 50 nm) shows no such splitting and all the other modes resemble pretty well with the Raman data obtained from unexposed flakes. Hence, the shift in the phonon frequency owing to electron doping can be completely ruled out as the possible origin of the splitting of the 129 cm$^{-1}$ mode in $MoTe_2$.

The effect of electron doping to stabilize the $T_d$ phase could also be verified by Raman measurements on exfoliated thin flakes 10% Re doped $MoTe_2$. Figures S15-S19 compare the Raman data obtained at 300 K and at 80 K for flakes of various thicknesses. We observe that a substantial change is present near the 129 cm$^{-1}$ mode (which splits at low temperature), while other modes show no substantial change. This again verifies that signatures of electron doping in 1T′ $MoTe_2$ are not manifested through frequency shifts of phonons but the stabilization of the $T_d$ phase.

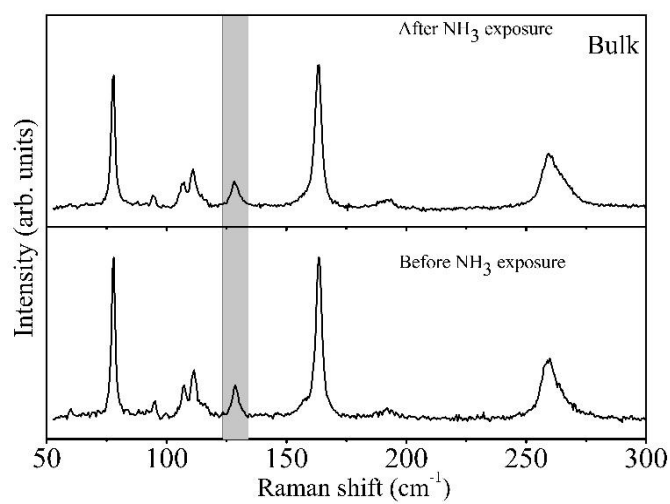

**Fig. S11. Raman spectra of bulk $MoTe_2$ (flake thickness ~ 50 nm) before and after exposure to ammonia vapour.**

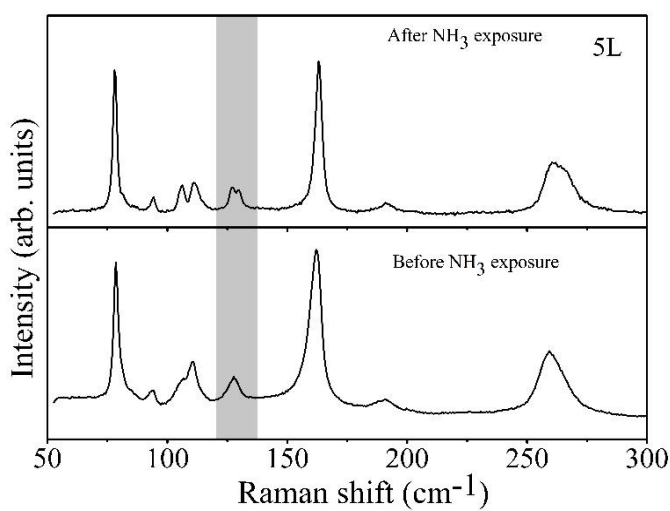

**Fig. S12. Raman spectra of 5-layered $MoTe_2$ flake before and after exposure to ammonia vapour.**

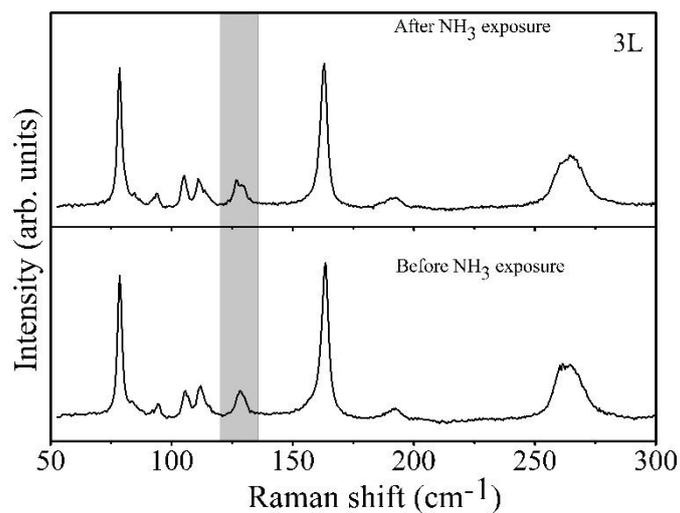

**Fig. S13. Raman spectra of 3-layered MoTe$_2$ flake before and after exposure to ammonia vapour.**

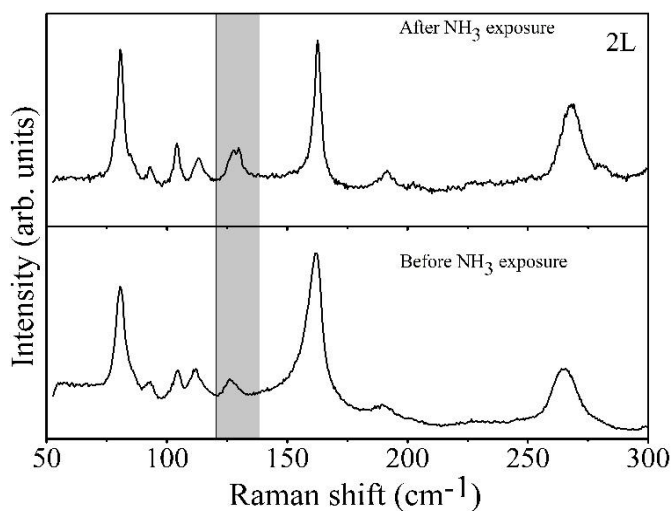

**Fig. S14. Raman spectra of 5-layered MoTe$_2$ flake before and after exposure to ammonia vapour.**

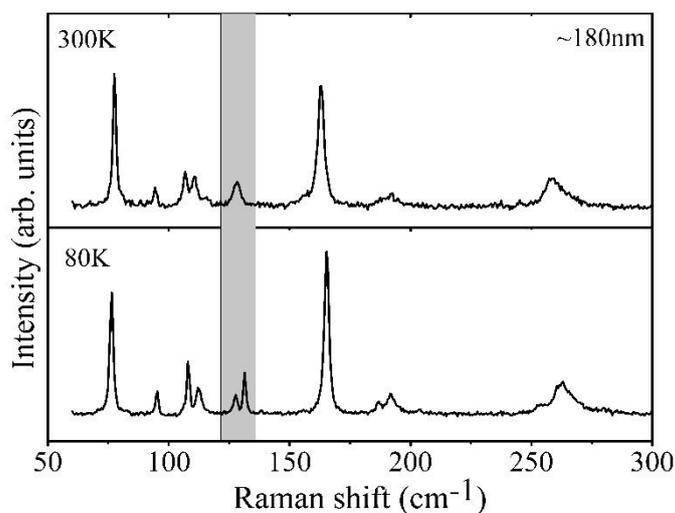

**Fig. S15. Raman spectra of 180 nm thick flake of 10% Re doped MoTe$_2$ at 300 K and 80 K, showing signature of phase transition.**

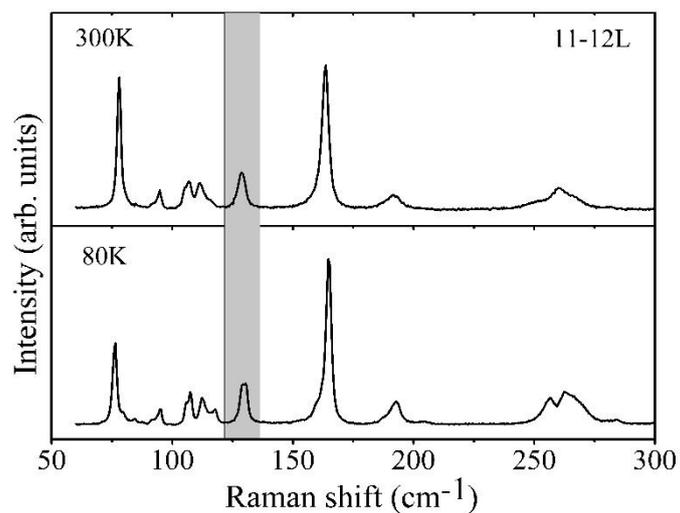

**Fig. S16. Raman spectra of 11-12 layers thick flake of 10% Re doped MoTe$_2$ at 300 K and 80 K, showing signature of phase transition.**

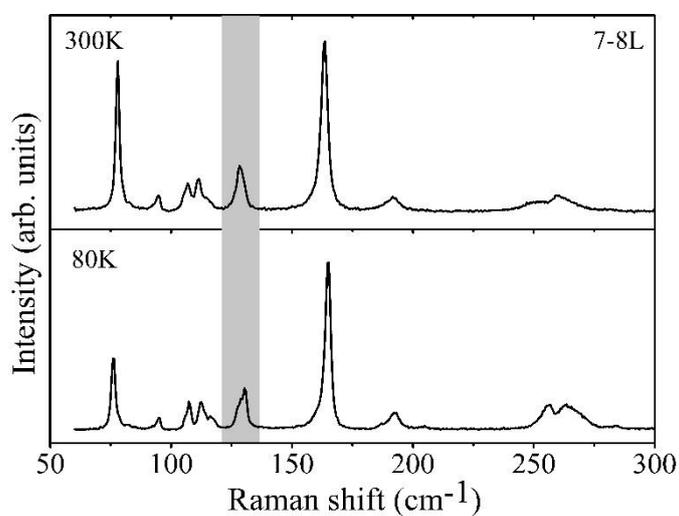

**Fig. S17. Raman spectra of 7-8 layers thick flake of 10% Re doped MoTe$_2$ at 300 K and 80 K, showing signature of phase transition.**

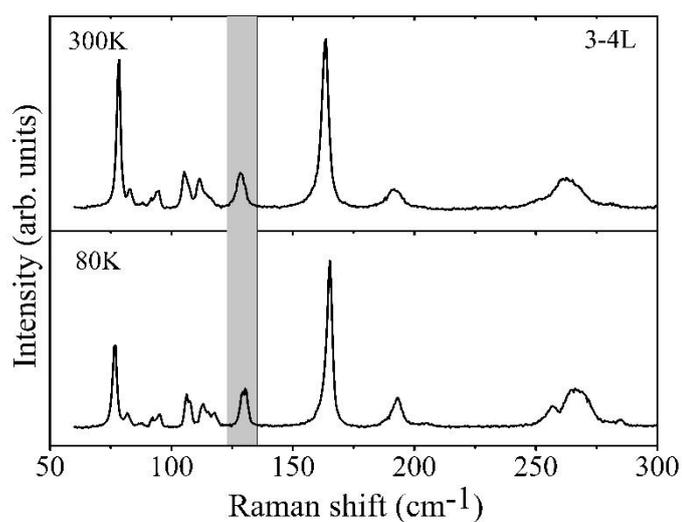

**Fig. S18. Raman spectra of 3-4 layers thick flake of 10% Re doped MoTe$_2$ at 300 K and 80 K, showing signature of phase transition.**

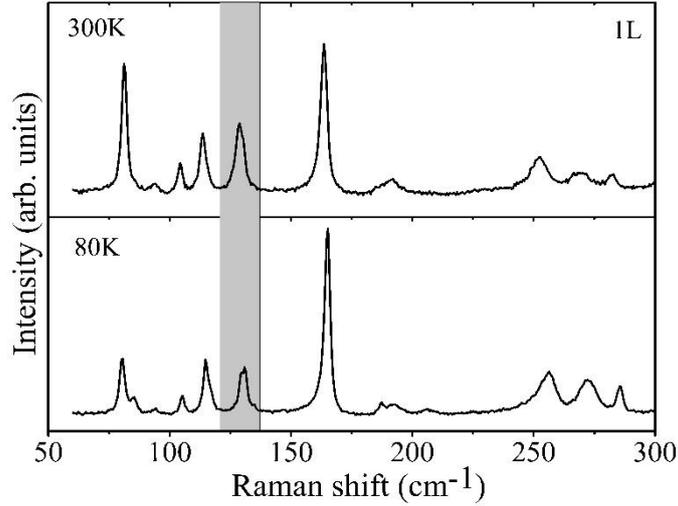

**Fig. S19.** Raman spectra of monolayer flake of 10% Re doped MoTe$_2$ at 300 K and 80 K, showing signature of phase transition.

## Supplementary Note 7:

*Study of electron-phonon coupling*

One of the most important results that we have obtained is the anomalous behaviour showcased by the phonon P1 (near 78 cm$^{-1}$) which shows a weak dependence on the thickness of the flake being studied. As discussed in Section 3.3 of main text and in Supplementary Note 4, the expected trend for phonon frequency and linewidth as a function of temperature is very well expressed by the anharmonicity theories that takes into account quasi-harmonic contributions and anharmonic contributions. However, a clear departure from the normal trend was observed in both the frequency and the linewidth of the mode P$_1$ for flakes of any arbitrary thickness (except the monolayer flake of pure MoTe$_2$). We attributed this anomalous behaviour to electron-phonon coupling in MoTe$_2$. Similar attribution was reported in graphene [8], and superconducting materials [9] previously. To further study the electron-phonon coupling as a function of flake thickness and to study the effect of electron/hole doping induced during the cleaving process, we have defined a parameter $\delta_{P1}(= \omega_{300\ K} - \omega_{80\ K})$. As also discussed in Section 3.5 of main text, the parameter $\delta_{P1}$ will have contributions from quasi-harmonic, anharmonic terms as well as electron-phonon coupling. However, a stronger contribution from the electron-phonon contribution will be manifested when $\delta_{P1}$ is positive. The Figures 7 and 9(b) show evolution of $\delta_{P1}$ as a function of thickness in pure and 10% Re doped MoTe$_2$ flakes. The corresponding Raman spectra (obtained at 300 K and 80 K) for the flakes of different thicknesses are shown in Figures S20 and S21. Figure S20 reveal the $\delta_{P1}$ values for the pure MoTe$_2$ flakes, while Figure S21 show the $\delta_{P1}$ values for the 10% Re doped MoTe$_2$ flakes.

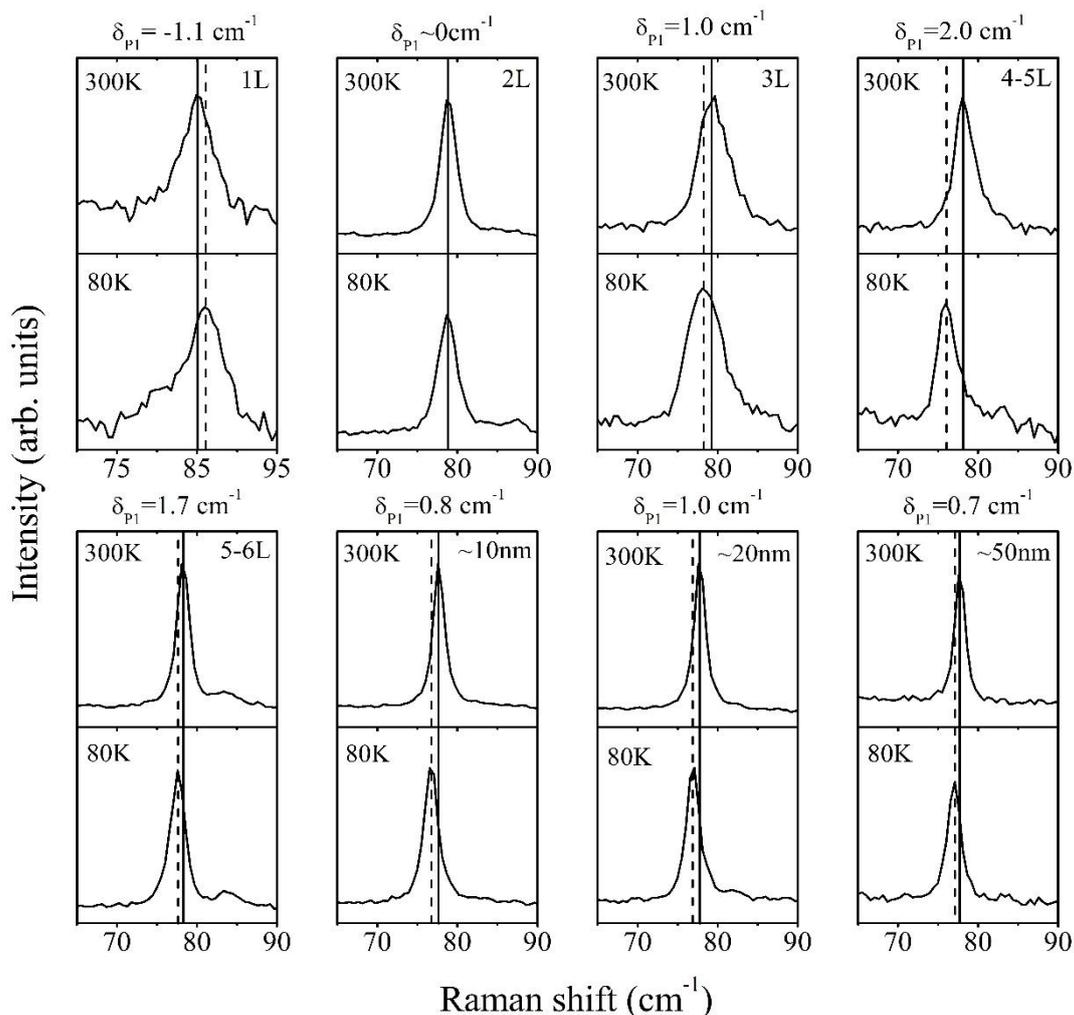

Fig. S20. Raman spectra of pure MoTe$_2$ flakes of various thicknesses obtained at 300 K and 80 K showing the anomalous shift in the P$_1$ mode. The solid (dashed) vertical lines represent the positions of the P$_1$ mode at 300 K (80 K). The corresponding measured values of δ$_{P1}$ are mentioned at the top of the figures.

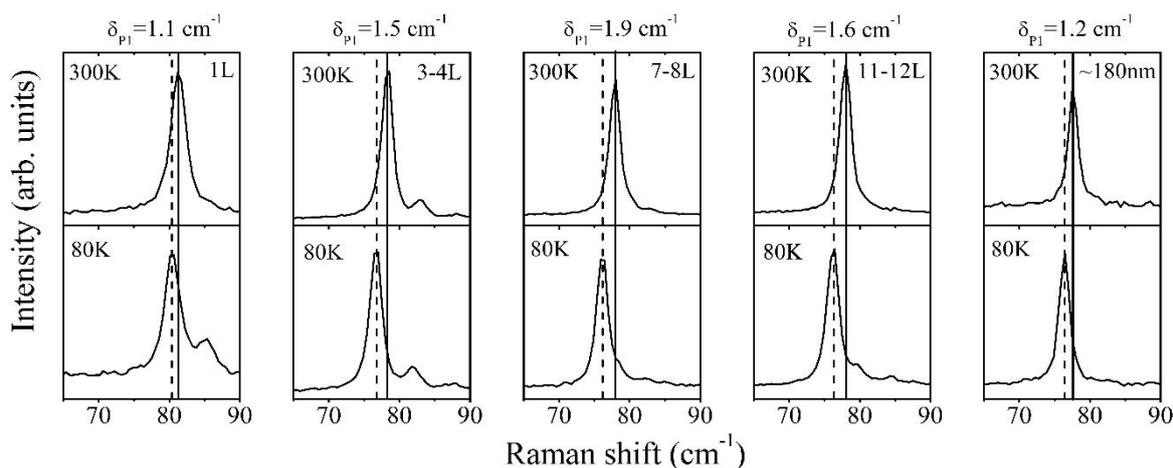

Fig. S21. Raman spectra of Mo$_{1-x}$Re$_x$Te$_2$ (x=0.1) flakes of various thicknesses obtained at 300 K and 80 K showing the anomalous shift in the P$_1$ mode. The solid (dashed) vertical lines represent the positions of the P$_1$ mode at 300 K (80 K). The corresponding measured values of δ$_{P1}$ are mentioned at the top of the figures.

## Supplementary Note 8:

*Competition between electron and hole doping in stabilizing 1T' or $T_d$ phase of MoTe$_2$*

Using ab-initio calculations, Kim *et al.* [10] predicted that charge (electron/hole) doping can influence the structural phase transition (1T′ ↔ $T_d$) in MoTe$_2$. A delicate balance between the electron and hole concentrations can switch from one phase to the other. Hole doping by ~ $10^{19}$ /cm$^3$ (refer to Fig. 5 of Reference [10]) would promote the stabilization of the 1T′ phase, while a charge neutral case or electron doping would lead to the stabilization of the $T_d$ phase. Considering a homogeneous and isotropic charge distribution, the required hole doping in 2-dimensions to stabilize the 1T′ phase is ~ $10^{12}$ /cm$^2$.

On the other hand, ARPES measurements by Pawlik *et al.* [11] demonstrated an electron doping in thinner flakes of MoTe$_2$ (exfoliated under controlled conditions) arising from Te-deficiencies and structural defects. They reported an inherent electron doping of ~ 0.19/f.u. that implies an electron density of ~ 0.76e per unit cell of MoTe$_2$. Taking a unit cell volume of ~ 307 Å$^3$ [10], one gets an electron concentration of ~$10^{20}$ /cm$^3$ that translates to a surface electron density of ~ $10^{13}$ /cm$^2$.

Based on the above qualitative estimates, it is clear that the delicate balance between the electron/hole concentrations ($10^{12}$ to $10^{13}$ /cm$^3$) can be tweaked by an external/internal charge doping with appropriate modification, thus resulting in the desired structural phase. In our experiments, the thin flakes of MoTe$_2$ were exfoliated under atmospheric conditions, naturally resulting in their exposure to atmospheric moisture and air. The moisture in the atmosphere gets adsorbed on to the surface of the 2D layer and forms a redox couple with oxygen which draws electrons from the 2D flake of MoTe$_2$, resulting in hole doping. This mechanism of hole doping in 2D flakes of graphene and TMDs due to exposure to atmosphere has been reported by various groups [12,13,14]. The extent of hole doping into 2D flakes by this technique has been reported to be ~ $10^{12}$/cm$^2$ [15]. Therefore, we may conclude that the extent of electron doping induced by structural defects and hole doping induced by atmospheric oxygen and moisture are comparable and are both close to the limit of electron/hole doping required to stabilize either the $T_d$ or 1T′ phase of MoTe$_2$, as predicted by Kim *et al.*[10]. Therefore, tuning the electron/hole concentrations can be easily achieved by appropriate modifications. Hole doping, which affects the top surface that is exposed to the atmosphere, is expected to show stronger effects on thinner flakes with an enhanced surface to volume ratio. This behaviour is reflected in our results which show stronger electron-phonon coupling in thicker flakes, but as we reach the limit of atomically thin flakes, the electron-phonon coupling is completely suppressed (Figure 7 in main text). To be noted that similar to our data, the electron doping induced by Te-vacancies (structural defects) in 2H-MoTe$_2$ thin flakes has been reported to be overshadowed by hole doping induced by atmospheric exposure [14].

On the other hand, for the 10 % Re doped MoTe$_2$ flakes, an additional electron doping ~ 0.1/f.u. is expected which would further increase the effective electron doping by a factor of ~1.5. Therefore, the effective electron doping in Re doped flakes always overshadows the surface hole doping due to atmospheric exposure. The increase in electron concentration in MoTe$_2$ due to Re doping has also been theoretically and experimentally studied by Manasi et al. [16]. This is again reflected in our results where the electron-phonon coupling prevails even in atomically thin monolayers of Re(10%)-MoTe$_2$ (Figure 9(b) in main text).